\documentclass[fleqn,usenatbib]{mnras}

\usepackage{newtxtext}
\usepackage[T1]{fontenc}
\usepackage{soul}
\usepackage{graphicx}
\usepackage{amsmath}
\usepackage{hyperref}
\usepackage{bm}
\usepackage[dvipsnames, table]{xcolor}
\usepackage{tikz}
\usepackage[normalem]{ulem}
\usepackage{orcidlink}
\usepackage{cellspace}
\usepackage{changepage} % in the preamble
\usepackage{multirow}
\usepackage{float} 
\usepackage{adjustbox}
\usepackage{tabularx}
\usepackage{fontawesome}
\usepackage{comment}
\usepackage{lineno}
\usepackage{gensymb} 
\usepackage{eso-pic}% http://ctan.org/pkg/eso-pic

\AddToShipoutPictureBG*{%
  \AtPageUpperLeft{%
    \hspace{0.25\paperwidth}%
    \raisebox{-3.5\baselineskip}{%
      \makebox[0pt][l]{\textnormal{DES-2022-0696}}
}}}%

\AddToShipoutPictureBG*{%
  \AtPageUpperLeft{%
    \hspace{0.25\paperwidth}%
    \raisebox{-4.5\baselineskip}{%
      \makebox[0pt][l]{\textnormal{FERMILAB-PUB-24-0200-PPD}}
}}}%

\DeclareRobustCommand{\VAN}[3]{#2}
\let\VANthebibliography\thebibliography
\def\thebibliography{\DeclareRobustCommand{\VAN}[3]{##3}\VANthebibliography}

\newcommand{\rvunits}{h^{-1}{\rm Mpc}}

\definecolor{Tomato}{HTML}{FF6347}

\title[DES Y3 voids $\times$ {\it Planck} CMB lensing]{The Gravitational Lensing Imprints of DES Y3 Superstructures on the CMB: A Matched Filtering Approach}

\author[Umut. E. Demirbozan et al.]{
\parbox{\textwidth}{
\large{U. Demirbozan,$^{1}$\thanks{E-mail: udemirbozan@ifae.es}
S.~Nadathur,$^{2}$\thanks{E-mail: seshadri.nadathur@port.ac.uk }
I.~Ferrero,$^{3}$
P.~Fosalba,$^{4}$
A.~Kovacs,$^{5,54}$
R.~Miquel,$^{1,6}$
C.~T.~Davies,$^{7}$
S.~Pandey,$^{8}$
M.~Adamow,$^{9}$
K.~Bechtol,$^{10}$
A.~Drlica-Wagner,$^{11}$
R.~A.~Gruendl,$^{12}$
W.~G.~Hartley,$^{13}$
A.~Pieres,$^{14}$
A.~J.~Ross,$^{15}$
E.~S.~Rykoff,$^{16}$
E.~Sheldon,$^{17}$
B.~Yanny,$^{18}$
T.~M.~C.~Abbott,$^{19}$
M.~Aguena,$^{20}$
S.~Allam,$^{21}$
O.~Alves,$^{22}$
D.~Bacon,$^{23}$
E.~Bertin,$^{24,25}$
S.~Bocquet,$^{26}$
D.~Brooks,$^{27}$
A.~Carnero~Rosell,$^{28,20}$
J.~Carretero,$^{29}$
R.~Cawthon,$^{30}$
L.~N.~da Costa,$^{20,28}$
M.~E.~S.~Pereira,$^{31}$
J.~De~Vicente,$^{32}$
S.~Desai,$^{33}$
P.~Doel,$^{27}$
S.~Everett,$^{34}$
B.~Flaugher,$^{21}$
D.~Friedel,$^{9}$
J.~Frieman,$^{21,11}$
M.~Gatti,$^{35}$
E.~Gaztanaga,$^{4,23}$
G.~Giannini,$^{29}$
G.~Gutierrez,$^{21}$
S.~R.~Hinton,$^{36}$
D.~L.~Hollowood,$^{37}$
D.~J.~James,$^{38}$
N.~Jeffrey,$^{27}$
K.~Kuehn,$^{39,40}$
O.~Lahav,$^{27}$
S.~Lee,$^{34}$
J.~L.~Marshall,$^{41}$
J. Mena-Fern{\'a}ndez,$^{42}$
J.~J.~Mohr,$^{43,26}$
J.~Myles,$^{44,45}$
R.~L.~C.~Ogando,$^{20,28}$
A.~A.~Plazas~Malag\'on,$^{46,47}$
A.~Roodman,$^{46,47}$
E.~Sanchez,$^{32}$
I.~Sevilla-Noarbe,$^{32}$
M.~Smith,$^{48}$
M.~Soares-Santos,$^{49,22}$
E.~Suchyta,$^{50}$
M.~E.~C.~Swanson,$^{9}$
G.~Tarle,$^{22}$
N.~Weaverdyck,$^{51,52}$
J.~Weller,$^{53,43}$
and 
P.~Wiseman$^{48}$
\begin{center} (  DES Collaboration) \end{center}
}
\parbox{\textwidth}{
\small \textit{The authors' affiliations are shown in Appendix~\ref{sec:affiliations}.}
}
}}

\date{Accepted XXX. Received YYY; in original form ZZZ}

\pubyear{2024}

\begin{document}
\label{firstpage}
\pagerange{\pageref{firstpage}}%--\pageref{lastpage}}
\maketitle
%Abstract of the paper

\begin{abstract}
Low density cosmic voids gravitationally lens the cosmic microwave background (CMB), leaving a negative imprint on the CMB convergence $\kappa$. This effect provides insight into the distribution of matter within voids, and can also be used to study the growth of structure. We measure this lensing imprint by cross-correlating the \emph{Planck} CMB lensing convergence map with voids identified in the Dark Energy Survey Year 3 data set, covering approximately 4,200 deg$^2$ of the sky. We use two distinct void-finding algorithms: a 2D void-finder which operates on the projected galaxy density field in thin redshift shells, and a new code, \texttt{Voxel}, which operates on the full 3D map of galaxy positions. We employ an optimal matched filtering method for cross-correlation, using the MICE N-body simulation both to establish the template for the matched filter and to calibrate detection significances. Using the DES Y3 photometric luminous red galaxy sample, we measure $A_\kappa$, the amplitude of the observed lensing signal relative to the simulation template, obtaining $A_\kappa = 1.03 \pm 0.22$ ($4.6\sigma$ significance) for \texttt{Voxel} and $A_\kappa = 1.02 \pm 0.17$ ($5.9\sigma$ significance) for 2D voids, both consistent with $\Lambda$CDM expectations. We additionally invert the 2D void-finding process to identify superclusters in the projected density field, for which we measure $A_\kappa = 0.87 \pm 0.15$ ($5.9\sigma$ significance). The leading source of noise in our measurements is \emph{Planck} noise, implying that data from the Atacama Cosmology Telescope (ACT), South Pole Telescope (SPT) and CMB-S4 will increase sensitivity and allow for more precise measurements.
\end{abstract}

% Select between one and six entries from the list of approved keywords.
% Don't make up new ones.


%%%%%%%%%%%%%%%%%%%%%%%%%%%%%%%%%%%%%%%%%%%%%%%%%%
%%%%%%%%%%%%%%%%% BODY OF PAPER %%%%%%%%%%%%%%%%%%
\section{Introduction}
Early large-scale galaxy surveys have revealed that the Universe's structure forms a cosmic web, featuring dense filaments and galaxy clusters alongside under-dense regions known as cosmic voids \citep{peebles80}. As observational data has expanded, these vast regions, primarily devoid of matter and dark matter, have gained heightened attention \citep[e.g., see][]{Hamaus:2016,raghunathan19,Nadathur2019,nadathur20,vielzeufy1,kovacs22isw,Woodfinden_2023}.

In particular, cosmic voids have proven to be useful tools for advancing cosmological studies. They offer a means to constrain the Neutrino mass sum \citep{lesg2006,lesg2013,villa2013,massara15,kreisch19,vielzeuf23}. Additionally, their abundance and density profiles aid in distinguishing Modified Gravity models from General Relativity (GR), with voids' ability to bypass screening mechanisms in high-density environments being particularly noteworthy \citep{li2012,Martinosheth2009,pisani19,khoury2004,vain1972}.

As a result, voids offer an environment that is sensitive to physics beyond the standard model of cosmology. They are sensitive to many effects, such as the growth rate of cosmic structure and redshift space distortions \citep{Hamaus:2016,Hamaus:2017a,Nadathur2020,woodfinden2022,Woodfinden_2023}, Alcock-Paczynski distortions \citep{Lavaux2012,Hamaus:2016,Nadathur2019}, weak gravitational lensing \citep{melchior14,clampitt15,20sanchez2d,Fang2019,raghunathan19}, baryon acoustic oscillations \citep{Kitaura2016}, and the integrated Sachs-Wolfe effect \citep{Sachs1967,Granett:2008,nadathurcrit16,Kovacs2019more,kovacs22isw}.

The large-scale structure of the Universe can influence the observed cosmic microwave background (CMB) and imprint secondary anisotropies. For instance, weak secondary CMB anisotropies, shaped by the evolving low-$z$ structure, provide key observational tests for dark energy. In particular, \citealt{Kovacs2019more, kovacs22isw} have shown an excess ISW signal from large voids that deviates from the predictions of the $\Lambda$CDM model. These findings have intensified interest in CMB lensing signal of voids, as this offers an alternative test of these secondary anisotropies.

Several recent works \citep[e.g.][]{schmittfull18,SO19,tanimura} have studied the cross-correlation between constructed weak CMB lensing convergence maps ($\kappa$) and large-scale structure. For example, it was shown that CMB lensing can be used to measure the masses of dark matter haloes \citep[initial detections include][]{madhavacheril15, baxter15,plancksz15} and to explore the non-linearities in structure formation through its correlation with cosmic filaments \citep{he18}.  However, unlike filaments and clusters, cosmic voids, as under-dense regions, cause an anti-lensing effect and imprint a negative $\kappa$ signal on the constructed CMB lensing map.

The CMB lensing imprints of voids have been measured by various authors \citep{cai17,raghunathan19, vielzeuf19, kovacsy3cmb, Hang2021} with detection significances ranging from $\sim3\sigma$ up to $\sim9\sigma$. However, \citet{kovacsy3cmb} found the amplitude of the lensing signal to be low, in moderate $\sim2\sigma$ tension with predictions from simulations, and \citet{Hang2021} also found hints of lower-than-expected lensing from voids. Given the other tensions in the late-time measurements of cosmological parameters with the predicted values from $\Lambda$CDM \citep{Riess2019,verde2019,heymanskids,snowmasstension}, any discrepancy in the CMB lensing signal from voids would be of great interest.

The stacked CMB lensing signal from voids is strongly influenced by specific void parameters. A typical void features a low-density core (with overdensity \(\delta < 0\)) which may be surrounded by a compensating marginally overdense region (\(\delta > 0\)). For most voids, the low central matter density gives rise to a de-lensing effect, characterized by \(\kappa < 0\) around the line of sight through the void center. However, depending on the relative amplitudes of the density fluctuations in the core and the surrounding overdensity and their physical extents, the convergence profile $\kappa(\theta)$ at angle $\theta$ from the void centre may differ, showing dips and rings of different widths \citep{Nadathur:2017a, Raghunathan2019}. This in turn can depend on the properties of the void-finding algorithm used for identification.

The objective of our study is to re-examine the CMB lensing imprint of voids in the DES Y3 data using a matched filtering method, as outlined in \citep{raghunathan19}, and to employ the \texttt{Voxel} void finder algorithm, which has not been previously utilized in DES. Moreover, we explore the variation of the CMB lensing signal with respect to redshift and the type of void finder employed. In contrast to prior studies by \citep{raghunathan19,Hang2021,kovacsy3cmb}, which each employed a single type of void finder, our study implements two different void finders to enhance the depth and range of our findings.

This paper is structured as follows: Section 2 explains our observed and simulated data sets. Section 3 outlines our methods for identifying voids and superclusters, as well as our matched filtering technique. Our primary observational findings are detailed in Section 4, while Section 5 offers a comprehensive discussion and concludes the study.
%%%%%%%%%%%%%%%%%%%%%%%%%%%%%%%%%%%%%%%%%%%%
%%%%%%%%%%%%%%%%%%%%%%%%%%%%%%%%%%%%%%%%%%%%
%%%%%%%%%%%%%%%%%%%%%%%%%%%%%%%%%%%%%%%%%%%%

%%%%%%%%%%%%%%%%%%%%%%%%%%%%%%%%%%%%%%%%%%%%
\section{DATA AND SIMULATIONS} \label{THE DATA SETS AND SIMULATIONS}
%%%%%%%%%%%%%%%%%%%%%%%%%%%%%%%%%%%%%%%%%%%%

\subsection{Observational Data}
\subsubsection{Dark Energy Survey Year-3 Data}
We identify cosmic voids using photometric redshift data from the first three years (Y3) of the Dark Energy Survey (DES). DES is a six-year sky survey, with Y3 covering approximately one-eighth of the sky (5000 deg$^2$) to a depth in the $i_{AB}$ band of less than 24, imaging around 300 million galaxies in five broadband filters ($grizY$) up to redshift $z=1.4$ \citep[for details see e.g.][]{DECam,morethanDE2016}. 

We use a luminous red galaxy sample from the DES Y3 dataset. This Red-sequence Matched-filter Galaxy Catalogue (\textit{redMaGiC}) \cite[see][for DES science verification (SV) test results]{Rozo2015} is a catalog of photometrically identified luminous red galaxies, with the red-sequence Matched-filter Probabilistic Percolation (redMaPPer) cluster finder algorithm \citep{Rykoff2014}. This algorithm has been used in many DES analyses \citep{Gruen2016,20sanchez2d,kovacs2017,Fang2019,vielzeufy1,kovacsy3cmb}.

The \textit{redMaGiC} galaxies offer the benefit of a low photo-$z$ error, approximately $\sigma_{z}/(1+z) \approx 0.013$. This error rate is half that of the \textit{MagLim} galaxy sample from DES Y3, which stands at $\sigma_{z}/(1+z) \approx 0.027$ \citep{Porredon_2022}. For comparison, we consider the study by \citep{kovacsy3cmb} which utilized an earlier version of \textit{redMaGiC}. Hence, we adopt the updated \textit{redMaGiC} v0.5.1 for this research.

The \textit{redMaGiC} algorithm produces various catalogs, distinguished by the densities and luminosities of the galaxies. Specifically, the High Density (HD) catalog maintains a consistent galaxy density, roughly \(\bar{n} \approx  10^{-3} h^3 \, \text{Mpc}^{-3}\) along the redshift range. Conversely, the High Luminosity (HL) catalog exhibits a galaxy density of \(\bar{n} \approx 4 \times 10^{-4} h^3 \, \text{Mpc}^{-3}\), which is considerably lower than its HD counterpart.

In our analysis, we focus on the redshift range 0.2 < $z$ < 0.8 and make use of the empirically constructed DES Y3 survey mask, which excludes contaminated pixels, mainly due to nearby stars. This low photo-$z$ error of \textit{redMaGiC} allows us to robustly identify void environments. The details of the clustering analysis of the DES Y3 \textit{redMaGiC} sample are documented in \citep{pandeyred}. It is also important to mention that this study identified certain systematic errors, to which our measurements are not sensitive, as we correlate them with an external LSS tracer, the CMB lensing map.

%%%%%%%%%%%%%%%%%%%%%%%%%%%%%%%%%%%%%%%%%%%%
%%%%%%%%%%%%%%%%%%%%%%%%%%%%%%%%%%%%%%%%%%%%

\subsubsection{Planck CMB Lensing Map}

We utilize the full-sky public Cosmic Microwave Background (CMB) lensing convergence ($\kappa$) maps from the Planck survey's 2018 data release \citep{planck18_lensing}.\footnote{Downloaded from \url{https://pla.esac.esa.int/\#cosmology}}

More specifically, we employ the \texttt{COM\_Lensing\_4096\_R3.00} map, which is reconstructed using a minimum-variance (MV) quadratic estimator \citep{hu02}. This estimator is based on a combination of foreground-cleaned {\it SMICA} \citep{planck16_smica} CMB temperature and polarization maps, with the mean field subtracted and a conservative mask applied to galaxy clusters to reduce contamination from thermal Sunyaev-Zel'dovich (tSZ) contributions. Throughout our analysis, we use $N_\mathrm{side} =512$ {\it HEALPix} maps \citep{gorski05}. We note that this $N_\mathrm{side} =512$ resolution is an appropriate choice considering the degree-scale imprints of voids.

The gravitational lensing of the CMB, occurs due to spatial variations in the gravitational potential field, $\Phi(r,\theta)$, as CMB photons traverse the Universe. The convergence can be described by the following equation:
\begin{equation}
\kappa(\theta) = \frac{3H_0^2\Omega_m}{2c^2} \int_0^{r_{\text{max}}} \frac{(r_{\text{max}} - r)r}{r_{\text{max}}} \delta(r, \theta) \, dr
\end{equation}
where $r$ is the comoving distance and $r_{\rm max}$ is the comoving distance to the last scattering surface of the CMB. This $\kappa$ is a dimensionless quantity and measures all the projected matter density up to the CMB surface and is weighted by a kernel for a given angular direction depending on the distance.
The gravitational potential is also related to the matter density fluctuation $\delta$ via the Poisson equation:
\begin{equation}
    \nabla^2\Phi=\frac{3H_0^2\Omega_m}{2a}\delta,
\end{equation}
where $\delta$ is the perturbation in the matter density and $a(t)$ is the dimensionless scale factor.
%%%%%%%%%%%%%%%%%%%%%%%%%%%%%%%%%%%%%%%%%%%%
%%%%%%%%%%%%%%%%%%%%%%%%%%%%%%%%%%%%%%%%%%%%
\subsection{Simulation - MICE CMB Lensing Map and \textit{redMaGiC} Tracers}

We utilize the publicly available MICE (Marenostrum Institut de Ciencias de l'Espai) simulation, which is an N-body light-cone extracted from the MICE Grand Challenge (MICE-GC). The MICE-GC contains approximately 70 billion dark-matter particles in a $(3h^{-1}\mathrm{Gpc})^3$ comoving volume. This simulation was created using the Marenostrum supercomputer at the Barcelona Supercomputing Center (BSC)\footnote{\textbf{\url{www.bsc.es}}}, running the \texttt{GADGET2} code \citep{springel2005}. For details on the creation of the MICE simulation, see \citet{fosalba2015a}, \citet{crocce2015}, and \citet{fosalba2015}.

The MICE simulation assumes a flat standard $\Lambda$CDM model with input fiducial parameters: $\Omega_m=0.25$, $\Omega_\Lambda=0.75$, $\Omega_b=0.044$, $n_s=0.95$, $\sigma_8=0.8$, and $h=0.7$, derived from the Five-Year Wilkinson Microwave Anisotropy Probe (WMAP) best fit results \citep{WMAPparam2009}.

In this study, we utilize the CMB lensing map from the MICE simulation, generated using the ``Onion Universe'' methodology as detailed in \cite{fosalba2008onion}. The validity of this lensing map was subsequently confirmed through auto- and cross-correlations with foreground MICE galaxy and dark matter particles (refer to \cite{fosalba2015} for an in-depth description of the map creation process). Initially, the MICE $\kappa$ map was provided with a \textit{HEALPix} pixel resolution of $N_{\rm side}=2048$. However, we downgraded the map to a lower resolution of $N_{\rm side}=512$. This adjustment significantly reduces the computational expense without causing a loss of much information, given that voids are degree-scale objects. We also downgrade the resolution of the {\it Planck} $\kappa$ map to $N_{\rm side}=512$  employed in our analysis.

We chose the \textit{redMaGiC} tracers on our mock galaxy catalog from MICE, maintaining consistency with the methodology utilized in the analysis of the observed DES Y3 data \cite{redmagic}. It is pertinent to note that the coverage of MICE dark matter halos is confined to an octant of the sky (5169.25 deg$^2$), which is larger than with the effective footprint of DES Y3 (4147.15 deg$^2$). This MICE-\textit{redMaGiC} mock galaxy catalog was constructed to match the number density of the DES Y3 \textit{redMaGiC} sample. It served as our tool to trace the distribution of galaxies on a large scale and to identify voids.

It is important to emphasize that the MICE cosmological parameters are relatively distant from the best-fit {\it Planck} cosmological parameters \cite{Planck2018_cosmo}. For example, the difference in the values of $\Omega_m$ and the Hubble constant $H_0$ can affect the amplitude of the lensing signal. However, we assume that variances in cosmological parameters, particularly $\Omega_m$, negligibly impact our lensing signal measurements. This postulation aligns with findings from \cite{vielzeufy1} and \cite{kovacsy3cmb}, who, utilizing the WebSky simulation \cite{webskysim}, showed minimal influence of $\Omega_m$  on the CMB lensing signal's amplitude.Additionally, our methodology encompasses a comprehensive error analysis, particularly addressing the MICE template uncertainties, ensuring the accuracy of our findings despite potential $\Omega_m$ discrepancies.

In this context, it is important to mention that, as detailed in \cite{vielzeuf23,nadathur19}, the parameter that seems to have the most significant impact on the determination of the matter content and the lensing convergence of voids is $\sigma_8$. The value of $\sigma_8$ in the MICE simulation, which is 0.8, is not far off from the best-fit Planck value of $\sigma_8 = 0.811 \pm 0.006$. Additionally, \citep{Nadathur2016} and \citep{nadathur15a} have identified the primary factors influencing the size and number of voids in any galaxy sample as the mean galaxy number density, the amplitude of galaxy clustering and the linear galaxy bias. Furthermore, $\sigma_8$ also affects void density profiles, especially close to void center \citep[e.g., see Figure 5 in][]{nadathur19}.

%%%%%%%%%%%%%%%%%%%%%%%%%%%%%%%%%%%%%%%%%%%%
%%%%%%%%%%%%%%%%%%%%%%%%%%%%%%%%%%%%%%%%%%%%
\section{METHOD} \label{METHOD}

Our approach aligns with the matched filtering technique detailed in \citep{nadathurcrit16,raghunathan19}. Notably, \cite{nadathurcrit16} evaluated the ISW imprint of voids, emphasizing that this technique avoids dependence on arbitrary choices of additional tuning parameters (such as the smoothing scale for Gaussian filtering of the CMB) that could introduce biases.

Utilizing the \texttt{Voxel} void parameters, \( \bar{\delta}_{g} \) and \( R_v \), we introduce a dimensionless parameter:
\begin{equation}
    \label{eq:lambda_v}
    \lambda_v \equiv \bar{\delta}_{g} \left(\frac{R_v}{1\ \text{Mpc}/h}\right)^{1.2}\,.
\end{equation}
This parameter, as empirically demonstrated by \cite{nadathur17}, exhibits a strong correlation with void density profiles and their macroscopic environments. As such, \( \lambda_v \) serves as a pertinent proxy for the gravitational potential associated with voids. Given this relationship, we expect notable variations in the lensing convergence profiles of voids based on their respective \( \lambda_v \) values.

%%%%%%%%%%%%%%%%%%%%%%%%%%%%%%%%%%%%%%%%%%%%%%%%%%%%%%
%%%%%%%%%%%%%%%%%%%%%%%%%%%%%%%%%%%%%%%%%%%%%%%%%%%%%%
\subsection{Void and Supercluster Finding }
\subsubsection{\texttt{Voxel} Voids}

The main goal of this research is to measure the CMB lensing signal from voids using two different void definitions: \texttt{Voxel} and 2D. The \texttt{Voxel} method is designed specifically for datasets with fragmented survey footprints, like DES Y3. One of the key benefits of \texttt{Voxel} is that it estimates the galaxy density field by computing number counts on a mesh, normalised by the counts of unclustered random points whose distribution accounts for the survey window function. This is the same method as used for estimating densities when computing power spectra: its use provides a natural way to account for variations in the survey selection function, and makes the \texttt{Voxel} algorithm better at handling gaps or fragmented survey masks than algorithms that employ Voronoi tessellations to estimate the density field. However, after the density field has been estimated, \texttt{Voxel} identifies voids using a watershed algorithm similar to that used by other, tesellation-based, algorithms \citep[e.g.]{neyrinck_zo, VIDE:2015,Nadathur:2019a}. .

The algorithm generates the mesh size, denoted as $N_{\text{mesh}}$, based on the tracer mean number density. The size of the mesh is set based on the condition that every cubic grid unit, known as a voxel, should possess a side length represented by \(a_{\text{vox}} = 0.5 \times \left( \frac{4\pi \bar{n}}{3} \right)^{-\frac{1}{3}}\). Here, \( \bar{n} \) stands for the estimated average density of galaxies.
The galaxy density field is then subsequently smoothed using a Gaussian filter of size $n_t^{-1/3}$, and local minima are identified across the voxels. Basins are formed around each local minimum, mirroring the process used in earlier \texttt{ZOBOV} algorithm \citep{neyrinck_zo}. The addition of adjacent voxels with increasing overdensity to the basin halts when the next voxel shows a lower density than its predecessor. Each resulting basin signifies a \texttt{Voxel} void.

For each identified \texttt{Voxel} void, we compute an average galaxy overdensity $\bar{\delta}_{g}$ and define an effective spherical radius $R_v$, which equates to the radius of a sphere with a volume equivalent to that of the void.

We generate the \texttt{Voxel} void catalogs for both DES Y3 and MICE simulation using the open-source \texttt{REVOLVER} void-finding code \cite{Nadathur2019}\footnote{Available at \url{https://github.com/seshnadathur/Revolver}}.

\begin{figure}
\begin{flushleft}
\begin{minipage}{0.95\columnwidth}
\includegraphics[width=\linewidth]{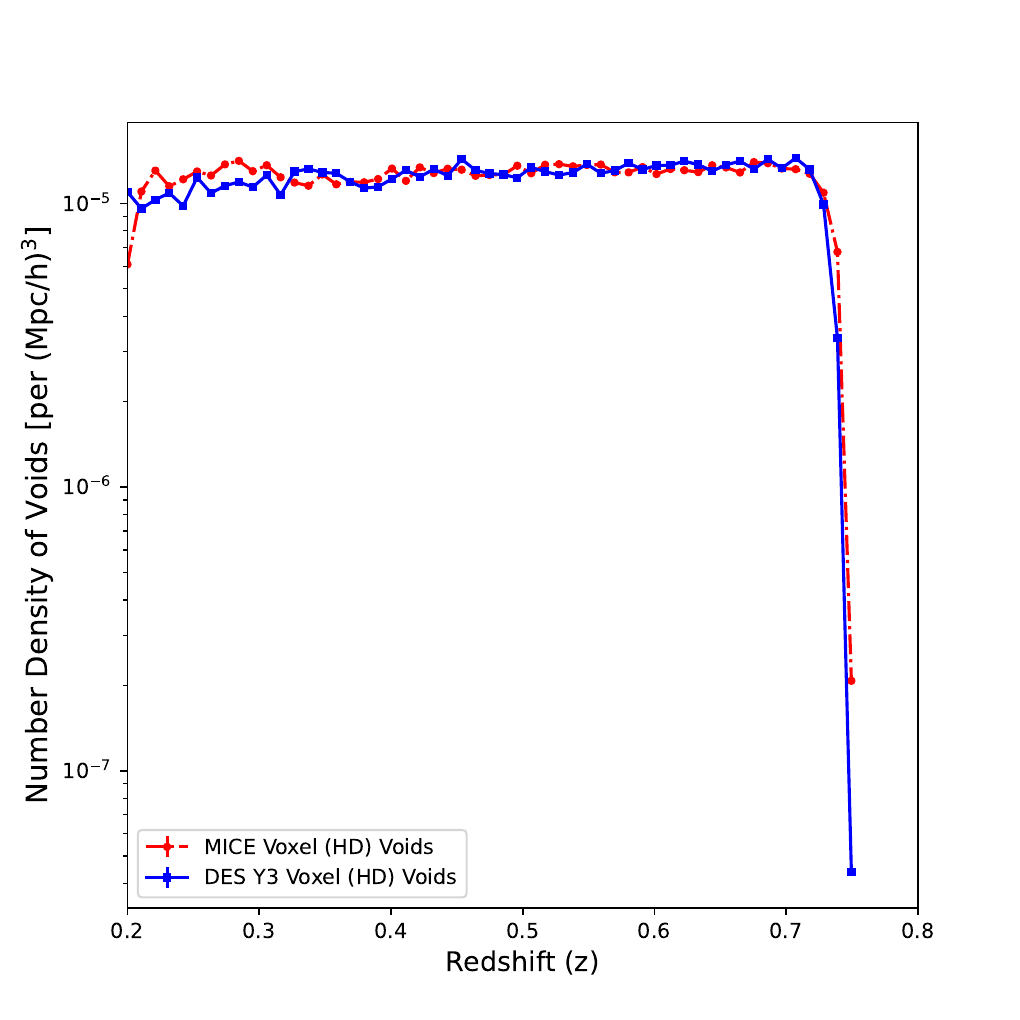}
\caption{Histogram illustrating the distribution of \texttt{Voxel} void number density per unit volume in both the MICE and DES Y3 HD catalogs.
}
\label{fig_voxel_void_size_histogram}
\end{minipage}
\end{flushleft}
\end{figure}

Void sizes in the MICE simulation range from $1.95\rvunits \le R_v\le 61.96\ \rvunits$, peaking around the median value $R_v=19.37\ \rvunits$. Meanwhile, for the DES, void sizes range from $2.04\rvunits \le R_v\le 59.41\ \rvunits$, peaking around the median value $R_v=19.98\ \rvunits$. Both the MICE and DES have a median void redshift of $z=0.57$. We perform a comparative analysis of the number density of \texttt{Voxel} voids identified in both the MICE and DES Y3 datasets per comoving volume $(\text{Mpc}/h)^3$, as shown in Figure \ref{fig_voxel_void_size_histogram}.

By applying the \texttt{Voxel} void finder to the updated \texttt{redMaGiC} galaxies, we obtained notable insights. Our preference for HD tracers of \texttt{redMaGiC} in identifying \texttt{Voxel} voids was influenced by their enhanced CMB lensing signal-to-noise ratio. Our tests confirmed that HD voids exhibit a higher S/N than HL voids, enhancing the efficacy of the matched filter method over other tracer densities. This outcome is primarily attributed to the reduced galaxy density in the HL.

Table~\ref{voxel_redshift_bin_size_table} outlines our findings: 44,426 voids in the MICE catalog and 33,427 voids in the DES Y3 catalog, both identified using High Density (HD) tracers. The disparity in void counts between MICE and DES Y3 is due to differences in effective sky coverage: MICE covers \(5169.25 \, \text{deg}^2\) while DES Y3 covers \(4143.17 \, \text{deg}^2\). Consequently, the total voxel void number densities for MICE and DES Y3 are \(8.59 \, \text{voids/deg}^2\) and \(8.07 \, \text{voids/deg}^2\), respectively. Although our primary interest lies in voids within the $0.2 < z < 0.8$ redshift range, data constraints restricted our \texttt{Voxel} void analysis in the MICE catalog to a maximum redshift of $z=0.75$.

%%%%%%%%%%%%%%%%%%%%%%%%%%%%%%%%%%%%%%%%%%%%%%%%
%%%%%%%%%%%%%%%%%%%%%%%%%%%%%%%%%%%%%%%%%%%%%%%%
\subsubsection{2D Voids}

 We also employ a 2D void finder, an algorithm initially developed for photometric redshift surveys. This algorithm identifies voids within tomographic redshift bins, which are slices of space at different redshift ranges \citep{20sanchez2d}. The 2D void finder has been employed in several DES studies \cite{vielzeufy1,Kovacs2019more,kovacsy3cmb}.

The 2D algorithm inspects possible minima in the galaxy density field, which has been projected and smoothed for each redshift slice. The void radius $R_{v}$ is defined when the density inside a thin annulus around the void center reaches the mean density of the redshift slice, increasing the annuli by $1\;h^{-1}$Mpc. While $R_{v}$ is typically measured in degrees and then converted to $\rvunits$ for consistency, we opt to use degrees in our matched filtering analysis.

Key parameters for the 2D void finder include the smoothing scale for the galaxy density maps, the redshift slice thickness, and the central minimum pixel density. We adopt a smoothing parameter of $\sigma = 10$ $\rvunits$, a central pixel density that is at least $30\%$ of the most underdense pixels in the redshift slice, and a slice thickness of $s \approx 100 \rvunits$. This results in 12610 and 10904 voids for MICE and DES Y3, respectively. 

To remove potentially spurious objects due to variations in photometric redshift, a measure of redshift derived from the photometric observations of an object, we apply a cut of $R_{v} > 20\rvunits$, following the precedent set in \cite{vielzeufy1} and \cite{kovacsy3cmb}. After this cut, the number of 2D voids reduces to 6295 for MICE2 and 5148 for DES Y3. 

\begin{figure}
\begin{flushleft}
\begin{minipage}{0.95\columnwidth}
\includegraphics[width=\linewidth]{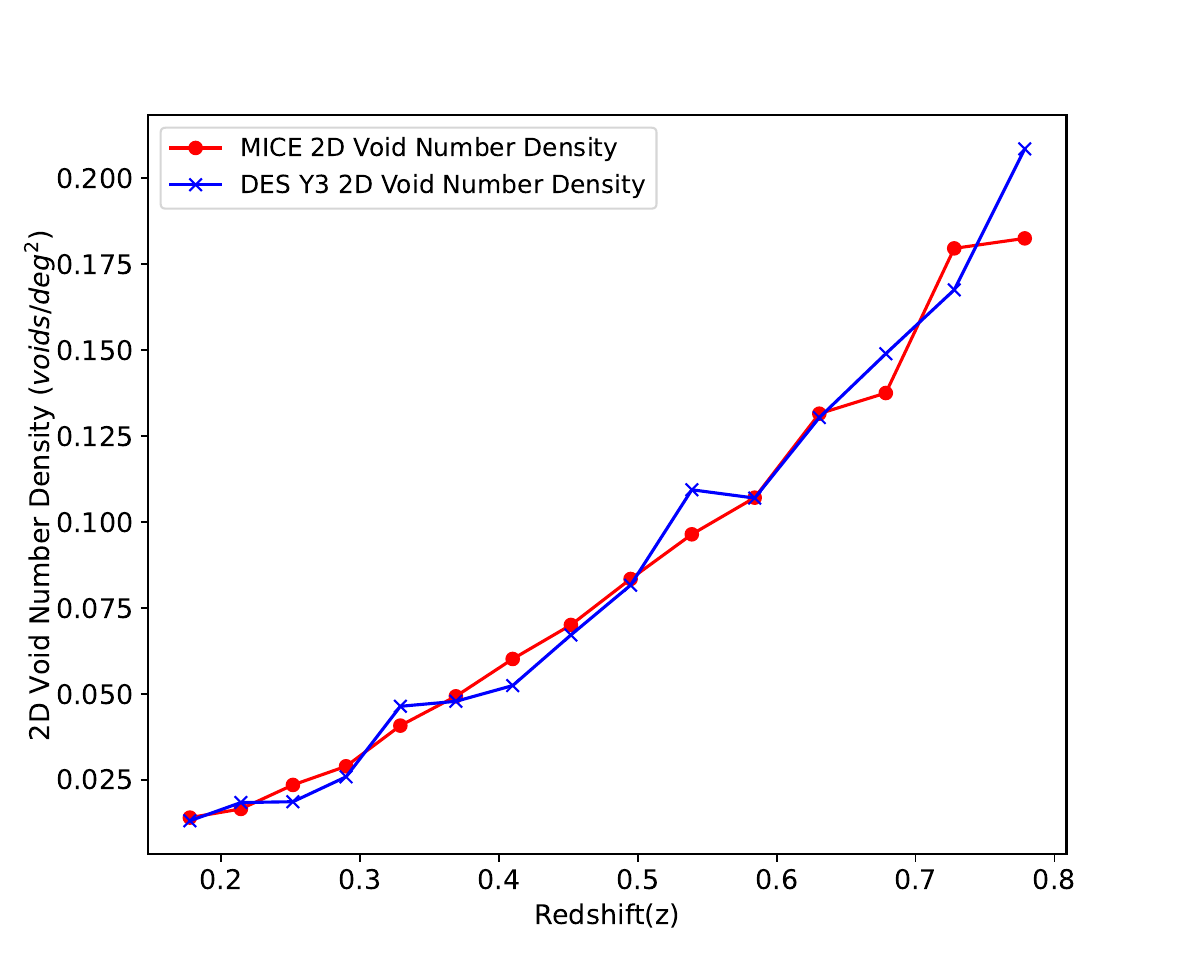}
\caption{Histogram illustrating the distribution of 2D void number density per unit volume as a function of tomographic redshift bins for the MICE and DES Y3.
}
\label{fig_2d_void_density}
\end{minipage}
\end{flushleft}
\end{figure}

 Normalizing by the effective area, the 2D void densities are approximately \(1.22 \, \text{voids/deg}^2\) for MICE2 and \(1.25 \, \text{voids/deg}^2\) for DES Y3 within the total redshift range. Just as with the \texttt{Voxel} voids, the 2D void densities of MICE and DES Y3 align closely. This implies that the discrepancies in absolute void count largely arise from the differences in effective areas: 5169 \( \text{deg}^2 \) for MICE2 and 4147 \( \text{deg}^2 \) for DESY3. Figure \ref{fig_2d_void_density} shows the 2D void number density per unit volume as a function of redshift bins. Specifically, we find that MICE (voids/$\text{deg}^2$) = 1.22, while DES Y3 (voids/$\text{deg}^2$) = 1.25, which indicate a close alignment between the two. For a detailed view, refer to Table \ref{tab:comparison_2d}.
%%%%%%%%%%%%%%%%%%%%%%%%%%%%%%%%%%%%%%%%
%%%%%%%%%%%%%%%%%%%%%%%%%%%%%%%%%%%%%%%%
%%%%%%%%%%%%%%%%%%%%%%%%%%%%%%%%%%%%%%%%%%%%%%%%
%%%%%%%%%%%%%%%%%%%%%%%%%%%%%%%%%%%%%%%%

For this analysis, we utilize the High Luminosity (HL) tracers from our updated \texttt{redMaGiC} v0.5.1 sample, this tracer density type choice is consistent with \cite{kovacsy3cmb} and has also been used with 2D voids in other DES studies \citep{kovacs22isw,kovacs2021eridanus}.  The study conducted by \cite{kovacsy3cmb} is particularly relevant to our analysis as it also examined DES Y3 2D voids defined on HL tracers and used the MICE simulation, similar to our approach. The key difference lies in their filtering and stacking method, which differs from ours. Therefore, it becomes intriguing to investigate the extent of the difference these distinct methods can cause, given the similarity in both the dataset and the simulation used. 
This choice was made with the aim of examining the effects of an alternative method, the matched filter, on our measurements. 

%%%%%%%%%%%%%%%%%%%%%%%%%%%%%%%%
%%%%%%%%%%%%%%%%%%%%%%%%%%%%%%%%

\subsubsection{2D Superclusters}

In order to supplement our analysis of 2D voids and provide further validation of our results, we have expanded our study to include 2D superclusters. To identify these superclusters, we applied the same 2D void finder to the tomographic galaxy density maps used in the previous void analysis, but with a twist: we inverted the density maps by multiplying them by $-1$. This operation effectively flips the galaxy density map, facilitating the detection of voids in the newly inverted galaxy density landscape. As such, the detected 'voids' in this inverted landscape correspond to superclusters in the original data. It is important to mention that our main reason for using this supercluster definition is to be consistent with \citep{kovacsy3cmb}. The superclusters found in this way represent overdense structures on a large scale within the distribution of galaxies, though they may not be bound by gravity. Similar to how voids are associated with extensive troughs in the density field, superclusters are linked to prominent peaks in this field \citep{nadathur14_void_catalogue_paper}. 

Using this method, we compiled a catalog of 2D superclusters, finding 5424 superclusters in the MICE simulation and 4432 superclusters in the DES Y3 dataset after applying a size cut of $R_{v} > 20\rvunits$. We then apply the same redshift binning as in 2D voids and we obtain the 2D supercluster numbers in Table \ref{tab:comparison_2d}.

%%%%%%%%%%%%%%%%%%%%%%%%%%%%%%%%%%%%%%%%%%%%%%%%%%%%%%%%%%%%%%%%%%%%%%%%%%%%%%%%%%%%%%%%%%%%%%%%%%%%%%%%%%%%%%%%%%%%%%%%%%%%%%

\subsection{Void Lensing in MICE Simulation}

\begin{figure*}
    \begin{minipage}{\textwidth}
        \centering
        \includegraphics[width=1.0\textwidth]{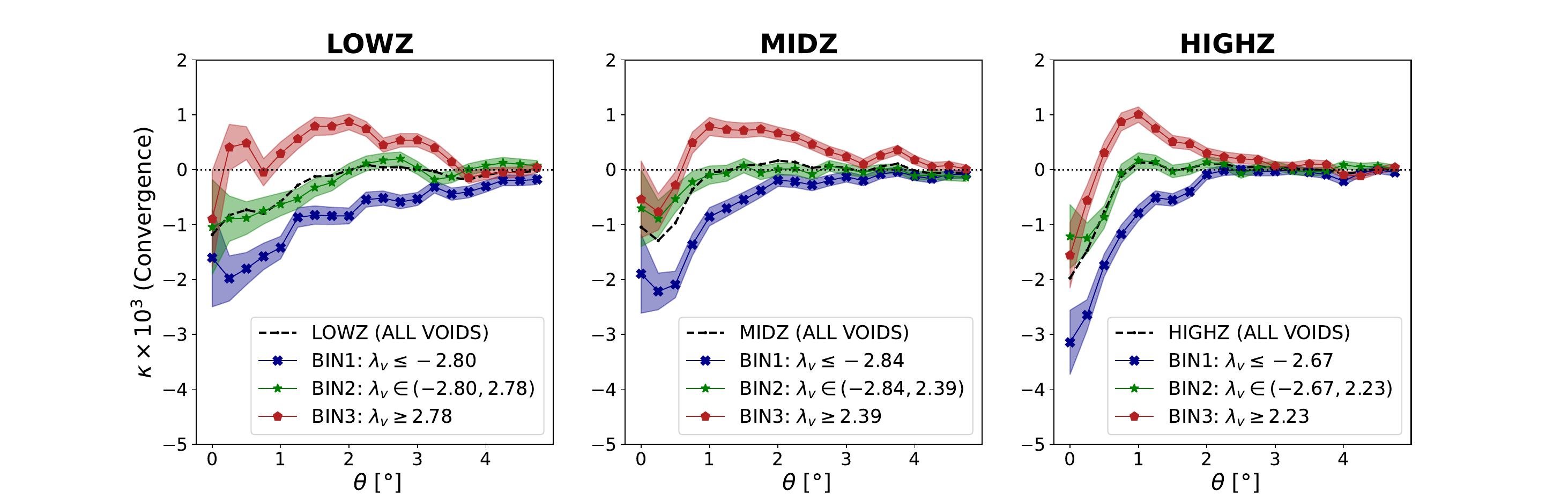}
        \caption{This figure presents the derived profiles from stacked images, each centered on identified \texttt{Voxel} voids within the MICE simulation. The data is segregated into three different redshift bins, and within each of these, there are three different $\lambda_v$ bins. The representation thus illustrates the behaviors and properties of these \texttt{Voxel} voids across three redshift bins and different values of $\lambda_v$. The shaded areas in the figure represent \(1\sigma\) uncertainty intervals, estimated through standard errors obtained from "void-by-void jackknife resampling." The dashed lines illustrate the aggregate result from all voids within the specified redshift category.}
        \label{fig:voxel_mice_templates}
    \end{minipage}
\end{figure*}

To begin, we inspect the void lensing imprints within the MICE simulation to calibrate the void lensing profiles $\kappa(\theta)$. This involves stacking 10$^{\circ}$ x 10$^{\circ}$ patches, derived from the full-sky MICE $\kappa$ map and centered on each void. These stacks reveal a negative $\kappa$ imprint (divergent lensing) at the center of the void, signifying the void's central underdensity. In contrast, a less conspicuous positive imprint encircles this, denoting matter overdensities at the outer periphery of the void. The measured radial profiles from these stacks can be seen in Figure \ref{fig:voxel_mice_templates}.

Unlike some studies \citep{kovacsy3cmb,cai17,Hotchkiss2015,vielzeufy1,Kovacs2019more,vielzeuf23} that adopt a re-scaling method, our measure of void lensing signals is in units of degrees $(\theta)$. This choice stems from our use of the matched filtering approach, akin to \citep{raghunathan19}, which necessitates that these simulation template void lensing profiles are defined as $\kappa(\theta)$. Importantly, by not re-scaling, we maintain uniformity in the noise power from the CMB lensing map across all void measurements. This is critical because the noise in the CMB lensing map is scale-dependent; thus, re-scaling might inadvertently mix different levels of noise power across various scales.

Additionally, \citep{nadathur17} has shown a stronger correlation between the void lensing signal and $\lambda_v$, as defined in Equation \ref{eq:lambda_v}—a function of void radius $R_v$ and void overdensity $\bar{\delta}_{g}$—than with the angular size of voids.

We organize the MICE void sample into three distinct redshift bins, labelled $\mathbf{LOWZ}$, $\mathbf{MIDZ}$, and $\mathbf{HIGHZ}$. This allows us to account for possible redshift-dependent variations in the CMB lensing imprints of voids.

\begin{table*}
\centering
\caption{This table presents the number of \texttt{Voxel} voids alongside the mean void size within each redshift bin for both MICE and DESY3 datasets with sky fraction being 0.125 and 0.100, respectively. For the detailed \texttt{Voxel} void size distribution across all 9 void bins see Figure \ref{fig:voxel_size_distributions} in the Appendix.}
\label{voxel_redshift_bin_size_table}
\begin{tabular}{lcccc}
\hline
& MICE Void Count & DES Y3 Void Count & MICE Mean Void Size & DES Y3 Mean Void Size \\
\hline
LOWZ (0.2 - 0.43) & 9298 & 6821 & 20.01 (Mpc/h) & 20.57 (Mpc/h) \\
MIDZ (0.43 - 0.59) & 14679 & 10861 & 20.22 (Mpc/h) & 20.54 (Mpc/h) \\
HIGHZ (0.59 - 0.75) & 20449 & 15027 & 19.58 (Mpc/h) & 20.02 (Mpc/h) \\
\hline
\end{tabular}
\end{table*}
Further granularity is achieved by subdividing each redshift bin into three separate $\lambda_v$ bins. Importantly, each $\lambda_v$ bin is populated with an approximately equal quantity of voids, a methodological decision that aligns with the approach used in \citep{raghunathan19}. By proceeding in this way, we derive a total of nine bins for our \texttt{Voxel} void sample.

As anticipated, the void lensing imprint strongly relies on the value of $\lambda_v$. More explicitly, voids with negative $\lambda_v$ values, such as those in bins 1 and 2, correlate with slightly larger, lower-density voids, exhibiting $\kappa \le 0$. On the other hand, voids with higher $\lambda_v$ values (as in bin 3) equate to "voids within voids" \cite{Sheth:2004}, predominantly smaller voids enveloped within large-scale underdensities, which present a positive $\kappa$ ring around the void boundary.

We also take into account the fact that the MICE footprint is significantly larger than the DES Y3 footprint and as a result, contains more voids. To assess the impact of this difference, we apply the DES Y3 mask to the MICE octant and identify the voids within this overlapping area. Subsequently, we carry out our stacking analysis using these voids. As expected, and in line with previous studies, we do not observe any substantial differences in the stacked profiles, despite the varying footprints. Consequently, we opt to use all the voids in the MICE simulation without implementing any footprint cut. This approach is consistent with the methodologies adopted by \citep{vielzeuf19,kovacsy3cmb}, providing further validation to our study.

%%%%%%%%%%%%%%%%%%%%%%%%%%%%%%%%%%%%%%%%%%%%
%%%%%%%%%%%%%%%%%%%%%%%%%%%%%%%%%%%%%%%%%%%%
%%%%%%%%%%%%%%%%%%%%%%%%%%%%%%%%%%%%%%%%%%%%
\subsection{Filtering the CMB lensing map}

When handling the CMB lensing map, we focus on lensing modes with $L \leq 2048$. This is because the lensing signals from cosmic voids are typically found on degree scales, and higher $L$ modes mostly consist of noise rather than useful signal. While we did consider the exclusion of the largest scale modes with $L < 8$ as was done in \cite{planck18_lensing}, we found that this had a negligible impact. As a result, we decided to exclude only the monopole and the dipole.

The key challenges in detecting the lensing signals are the lensing reconstruction noise present in the \textit{Planck} $\kappa$ map and the additional $\kappa$ values contributed by unrelated structures in our line of sight. These noise sources are about ten times larger than the lensing signal from a single void, rendering it essentially invisible. Hence, our strategy involves stacking these signals together and dividing the voids based on their $\lambda_v$ values. We then apply an optimized 'matched filter' to the $\kappa$ map before stacking. This filter, derived from templates in the MICE simulations, combines with the original map in a way that highlights the parts of the map that match the filter, making the lensing signals stand out more.

The matched filter transforms the original lensing map, represented as $\kappa$, into a 'filtered' version, or $\kappa^\mathrm{MF}$. This transformation involves convolving the filter with the original map, effectively amplifying the signals that match the filter. We created these filtered maps for each of nine different void 'bins'.

One of the key benefits of this matched filter technique is its neutrality - it is applied to the whole map uniformly so, it  doesn't unfairly favor certain parts of the map over others. It also reduces the variability, or 'noise', at the locations of voids to the lowest possible level. 

By comparing these measurements across all voids stacked together, we can see the benefits of our approach to split the data into bins based on $\lambda_v$ values. We also find that the lensing signals are most easily detected in the two extreme $\lambda_v$ bins rather than the middle bin. 

To construct the matched filters, we initiate by defining the convergence field at a location $\boldsymbol{\theta}$, originating from the position $\boldsymbol{\theta}_0$ which corresponds to the $\kappa$ value at the center of a void as illustrated in Figure \ref{fig:voxel_mice_templates}.
This is expressed as:
\begin{equation}
    \label{eq:kappa0}
    \kappa(\theta) = \kappa_{\text{template}}(|\theta - \theta_0|; \lambda_v) + p(\theta)
\end{equation}
In this context, $p$ is symbolic of the noise component in all the $\kappa$ maps, excluding the void lensing signal itself. Meanwhile, the $\kappa_{\text{template}}$ symbol represents the corresponding void lensing template profile, which is derived from the stacked images from the MICE.

It is possible to separate this template profile as:
%\begin{equation}
\begin{align} 
\kappa_{\text{template}}(\theta; \lambda_v) &= \kappa_0(\lambda_v)k(\theta; \lambda_v) \\
&= \kappa_0(\lambda_v)\sum_{L=0}^{\infty}k_{L0}(\lambda_v)Y_L^0 (\cos \theta),
\end{align}
%\end{equation}

In the given equations, $\kappa_0(\lambda_v)\equiv\kappa_{\text{template}}(0;\lambda_v)$ is what we refer to as the 'amplitude term'. It informs us about the maximum intensity of the template profile at the void center. Furthermore, $k(\theta)$, which we call a 'shape function', standardizes the shape of the template profile based on the coefficients of its spherical harmonic, $k_{L0}$. To derive the $k_{L0}$ coefficients, an interpolated univariate spline was employed to construct a HEALPix template map from the $k(\theta)$ measurements. This map’s pixel values, calculated from spherical coordinates, were then normalized relative to the value of $\kappa_0$.

By taking these definitions into account, and making the assumption that the noise component is uniform, shows no directional bias, and averages out to zero, we are in a position to calculate the spherical harmonic coefficients for the ideal matched filter:
\begin{align}
\Psi_{\text{L0}}^{MF}(\lambda_v) = \frac{\alpha k_{L0}{(\lambda_v)}}{C_{L}^{N_{\text{tot}}}}\,
\end{align}
with $ \alpha$  defined as
\begin{align}
\label{eq:alpha}
\alpha^{-1} \equiv \sum_{L=0}^{\infty} \frac{(k_{L0}{(\lambda_v)})^2}{C_{L}^{N_{\text{tot}}}}\,,
\end{align}
where
\begin{equation}
C_{L}^{N_{\text{tot}}} = C_{L}^{\kappa\kappa} + N_L^{\kappa\kappa}\,,
\end{equation}
is the total power spectrum, and \( {C_{L}^{\kappa\kappa}} \) and \( N_{L}^{\kappa\kappa} \) are the lensing and noise power spectra, respectively, for the Planck lensing map. 

\begin{figure*}
    \begin{minipage}{1.0\textwidth}
        \centering
        \includegraphics[width=1.0\textwidth]{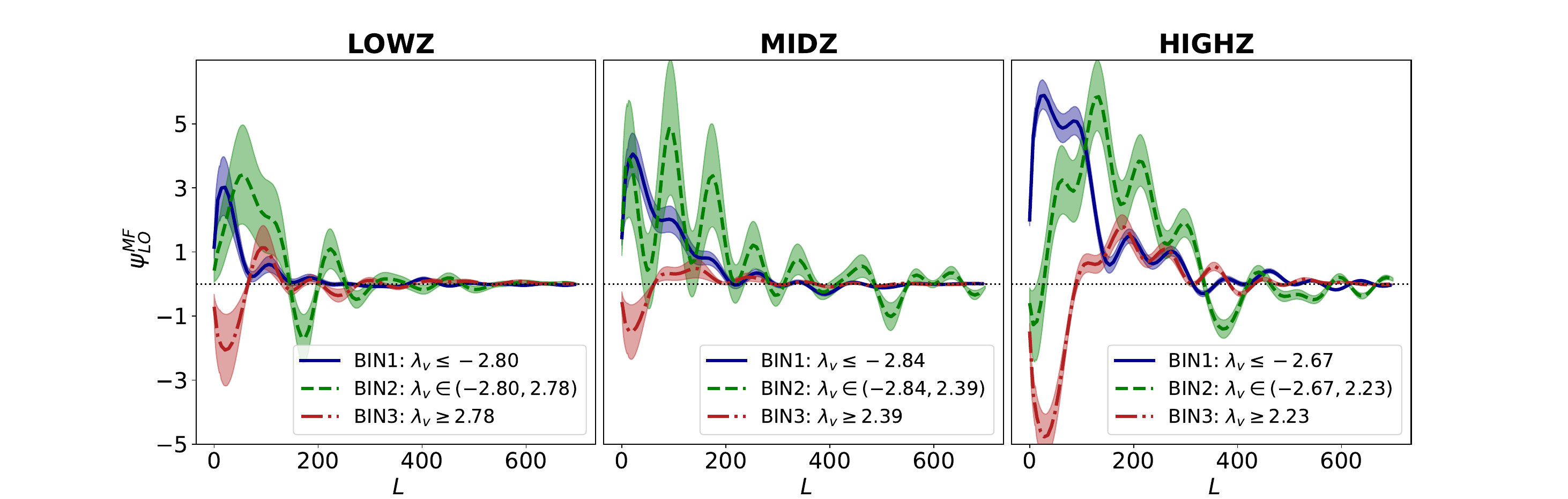}
        \caption{This figure shows the optimally determined spherical harmonic coefficients, known as the kernel of the matched filters \(\psi_{LO}^{MF}\), derived from the MICE template profiles shown in Figure \ref{fig:voxel_mice_templates}. The shaded regions represent uncertainty estimates calculated from 1000 synthetic datasets using jackknife standard errors, as detailed in Equation \ref{eq:jackknife}. In our analysis, we confine the application of spherical harmonics to \(L = 700\), a convention based on the observation that the power of these coefficients approaches zero beyond this threshold.}
        \label{fig:voxel_mice_mf}
    \end{minipage}
\end{figure*}

The ideal matched filters for each void bin, constructed using previously acquired template profiles, are displayed in Figure \ref{fig:voxel_mice_mf}. We calculate the sum in Eq.~\ref{eq:alpha} up to $L=700$, as the spherical harmonic coefficients for higher values of $L$ rapidly approach zero, as seen in the figure. We have also transferred the jackknife errors from the template profiles into the matched filters.

An important takeaway from Figure \ref{fig:voxel_mice_templates} is that the templates for the $\lambda_v$ bin 1 in every redshift category don't undergo a sign change, which suggests that the corresponding matched filters for these bins also stay the same. For all the remaining bins, a conspicuous crossover point is observed, resulting in filter profiles that are either partly or entirely compensated. In this context, a "compensated" profile refers to a matched filter profile where the areas under the curve with positive and negative values of kappa cancel each other out, leading to a net zero integral over the profile. This compensation phenomenon actually reflects a balance between the regions of under-density and over-density within the voids and superclusters, as captured by the lensing signal.

%%$%%%%%%%%%%%%%%%%%%%$
%%$%%%%%%%%%%%%%%%%%%%$
%%$%%%%%%%%%%%%%%%%%%%$
%%$%%%%%%%%%%%%%%%%%%%$
%%$%%%%%%%%%%%%%%%%%%%$
For each bin defined by the matched filter, based on $\lambda_v$, the lensing map which has been filtered, denoted by $\kappa_{\text{MF}}$, can be viewed as a convolution of the filter with the original map. Formally,

\begin{equation}
\kappa^{\text{MF}}(\eta) = \int d\Omega \kappa(\theta) \Psi^{\text{MF}} (|\theta - \eta|),
\end{equation}

This can also be transcribed into the spherical harmonic space \citep{Schaefer:2006} as follows,
\begin{equation}
\kappa_{\text{LM}}^{MF} = \sqrt{\frac{4\pi}{2L + 1}} \kappa_{LM} \Psi_{\text{L0}}^{MF}.
\end{equation}
The matched filter is designed to ensure the expected value of the filtered field at the void locations is,
\begin{equation}
\langle \kappa^{\text{MF}}(0; \lambda_v) \rangle = \kappa_0(\lambda_v),
\end{equation}
such that the filter is unbiased. This is the $\kappa_{0}$ measured at the void center. This matched filter also minimizes the variance of the filtered field at this location, given by
\begin{equation}
\sigma_{\text{MF}}^2(0; \lambda_v) = \sum_{L=0}^{\infty} C_{L}^{N_{\text{tot}}} |\Psi_{\text{L0}}^{MF}|^2\,. 
\end{equation}
The maximum detection level of the optimal matched filter, for a single isolated void, can be quantified as \citep{McEwen:2008},
\begin{equation}
\Gamma_{single}(\lambda_v) \equiv \frac{\langle \kappa_{\text{MF}}(0; \lambda_v) \rangle}{\sigma_{\text{MF}}(0; \lambda_v)} = \alpha^{-1/2} \kappa_0(\lambda_v)\,.
\end{equation}

In summary, matched filtering is applied to optimize the signal-to-noise ratio. This technique emphasizes the expected lensing signal in specific regions of the lensing map, following a predetermined template, while reducing non-relevant areas. The lensing signal, as represented by the void center pixel in the filtered map, enables a statistically optimized quantification of the lensing effect, which is suitable for stacking.

As observed in Figure \ref{fig:voxel_mice_mf}, it is important to note that the power of the filter primarily lies in the $L$ modes less than 500. This makes the final central pixel of voids susceptible to smaller-scale noise as we anticipate that high $L$ modes are dominated by noise. This does not compromise its effectiveness but rather highlights its specificity. Given that cosmic voids are large-scale structures, it is reasonable that these $L$ modes capture the most significant lensing information. Therefore, the efficiency of the matched filter is not predicated on capturing all possible information, but on maximizing the detectability of the specific lensing signal, assuming that the MICE template accurately characterizes this signal.

%%%%%%%%%%%%%%%%%%%%%%%%%%%%%%%%%%%%%%%%%%%%%%%%%
%%%%%%%%%%%%%%%%%%%%%%%%%%%%%%%%%%%%%%%%%%%%%%%%%
%%%%%%%%%%%%%%%%%%%%%%%%%%%%%%%%%%%%%%%%%%%%%%%%%
%%%%%%%%%%%%%%%%%%%%%%%%%%%%%%%%%%%%%%%%%%%%%%%%%
\subsection{Error Estimation} \label{Error Estimation}

We use the jackknife method to estimate the uncertainties in the MICE void lensing templates and their corresponding matched filters. This approach contrasts with the method used by \citet{kovacsy3cmb} (hereafter referred to as K22), which assumes that errors in MICE are negligible. We set $N_{jk} = N_{\text{void}}$ for each void bin in the MICE simulation. This process results in a jackknife sample of size $N_{\text{void}}$. The standard errors of the MICE templates, as used in Figure \ref{fig:voxel_mice_templates}, are given by:
\begin{equation}
\label{eq:jackknife}
\sigma_{\text{JK}} = \sqrt{\frac{N_{\text{void}} - 1}{N_{\text{void}}} \sum_{i=1}^{N_{\text{void}}} (\theta_i - \overline{\theta})^2}
\end{equation}
where $\theta_i$ are the individual jackknife samples, $\overline{\theta}$ is the mean of the jackknife estimates, and $N_{\text{void}}$ is the number of voids in that bin.

We then estimate the covariance matrix from these "void by void" jackknife samples by using \citep{Mohammad_2022}:
\begin{equation}
C_{ij}^{(jk)} = \frac{N_{\text{void}} - 1}{N_{\text{void}}} \sum_{k=1}^{N_{\text{void}}} (\theta_{k_i} - \bar{\theta_i}) (\theta_{k_j} - \bar{\theta_j})
\end{equation}
where the index $k$ represents the jackknife samples and $\theta_{k_i}$ and $\theta_{j_i}$ are the corresponding $\theta$ measurements in the $i$th bin, as shown in Figure \ref{fig:voxel_mice_templates}.

After this, a synthetic data vector is created as:
\begin{equation}
    \label{eq:synthetic}
    \xi_{\text{mock}} = LZ + \xi_{\text{th}},
\end{equation}
where \( LL^T = C_{ij} \), L is obtained by a Cholesky decomposition of the covariance matrix \( C_{\text{ij}} \), and \( Z \) is a vector of independent standard normal random variables ( $\mu = 0$ and $\sigma= 1$), and $\xi_{\text{th}}$ is the mean signal template for the corresponding bin. In this way, we conserve the structure of the covariance and add random Gaussian noise by using \(Z\), which makes our template error analysis more robust unlike previous CMB x LSS studies \citep{Kovacs2019, kovacs2021eridanus, vielzeuf19,gisela}

We then obtained $N_{\text{void}}$ synthetic data vectors using Equation \ref{eq:synthetic} for each void bin and calculated the mean profile of these. This process was repeated $N = 1000$ times to obtain 1000 mean profiles. We applied our matched filter methodology to obtain the spherical harmonic coefficients based on these 1000 mean profiles, as shown in Fig \ref{fig:voxel_mice_mf}. 

Subsequently, we perform a convolution of these matched filters with 1000 randomly generated MICE $\kappa$ maps, employing the synfast function from the Healpy library. This operation introduces an additional layer of randomness, derived from both the $\kappa$ map and the coefficients of the matched filters, enabling us to observe the impact of errors in the MICE templates. Otherwise, we could have just used the mean template profile to convolve with 1000 randomly generated MICE $\kappa$ maps, but we wanted to observe the effect of the uncertainty in the template profiles.  

We further tested our error analysis by applying jackknife resampling to different sub-volumes and using various groupings with different $N_{jk}$ values, such as 50, 64, and 100, instead of treating each void as an individual jackknife sub-sample. Our findings indicated that when implementing Equation \ref{eq:synthetic}, the standard deviation of the resulting templates remained consistent across these different groupings. This consistency reinforced our decision to employ the "void by void" jackknife approach for our final error analysis. This error analysis was repeated for 2D superclusters as well.

Intriguingly, we find that the final stacked $\kappa$ value emanating from the DES Y3 void centers remains unaffected by the randomness in the spherical harmonics coefficients of the matched filters. This can likely be attributed to the fact that the influence of this jackknife randomness is minimal at smaller scales, such as the center pixels of voids. We, therefore, conclude that our measurements are predominantly dominated by the noise in the Planck $\kappa$ map.

We first extract $\kappa_{0}$ values from the randomly generated and matched filtered Planck-like convergence maps at the location of the central pixels of each bin of DES Y3 voids. We then average out these values to calculate $\kappa_{0}$. This process, involving 1000 random instances of $\kappa_{0}$, allows us to construct a covariance matrix for our measurements. 

Next, we determine the $\kappa_{0}$ values at the central pixels of DES Y3 void positions for each bin on the corresponding matched filtered Planck map. These measured $\kappa_{0}$ values are subsequently compared with the corresponding MICE $\kappa_{0}$ values. Figure \ref{fig:final_result_void} illustrates the measured $\kappa_{0}$ values for each void bin, encompassing both \texttt{Voxel} and 2D voids. The standard errors incorporated into the plot correspond to the diagonal entries of our established covariance matrices.

To measure our CMB lensing detection significance, we use the $\chi^2$ minimization technique as in \citep{kovacsy3cmb, vielzeuf19}. We make use of the following equation;

\begin{equation}
    \label{eq:chi2}
\chi^2 = \sum_{l,m} \left(\kappa^{\text{DES}}_l - A_\kappa \cdot \kappa^{\text{MICE}}_l\right) C^{-1}_{lm} \left(\kappa^{\text{DES}}_m - A_\kappa \cdot \kappa^{\text{MICE}}_m\right)
\end{equation}

where $\kappa_l$ denotes the mean CMB lensing signal within $\theta$ bin $l$, and $C$ represents the related covariance matrix. We checked to identify the best-fitting amplitude, represented as $A_\kappa \pm \sigma_{A_\kappa}$, by constraining the shape of the stacked convergence profile from DES Y3 x Planck to align with the shape calibrated from the MICE simulation. Moreover, in our matched filtering methodology, we apply the Percival correction factor when inverting covariance matrices as described in \citep{Percival_2021}. Conversely, for the covariance matrix derived from the template fitting methodology of K22, we employ the Anderson-Hartlap factor $h = (n_{\text{randoms}} - 1) / (n_{\text{randoms}} - n_{\text{data points}} - 2)$ before inverting the covariance matrix \citep{hartlap2007}. This approach is adopted to maintain consistency with the methods utilized by K22. We then obtain the final results using the template fitting methodology of K22 as depicted in Figure \ref{fig:rescaled_results}.
%%%%%%%%%%%%%%%%%%%%%%%%%%%%%%%%%%%%%%%%%%%%%%%%%%%%%%%%%%%%%%%
%%%%%%%%%%%%%%%%%%%%%%%%%%%%%%%%%%%%%%%%%%%%%%%%%%%%%%%%%%%%%%%

\section{RESULTS}

\begin{figure*}
    \begin{minipage}{\textwidth}
    \centering
    \includegraphics[width=1.05\textwidth]{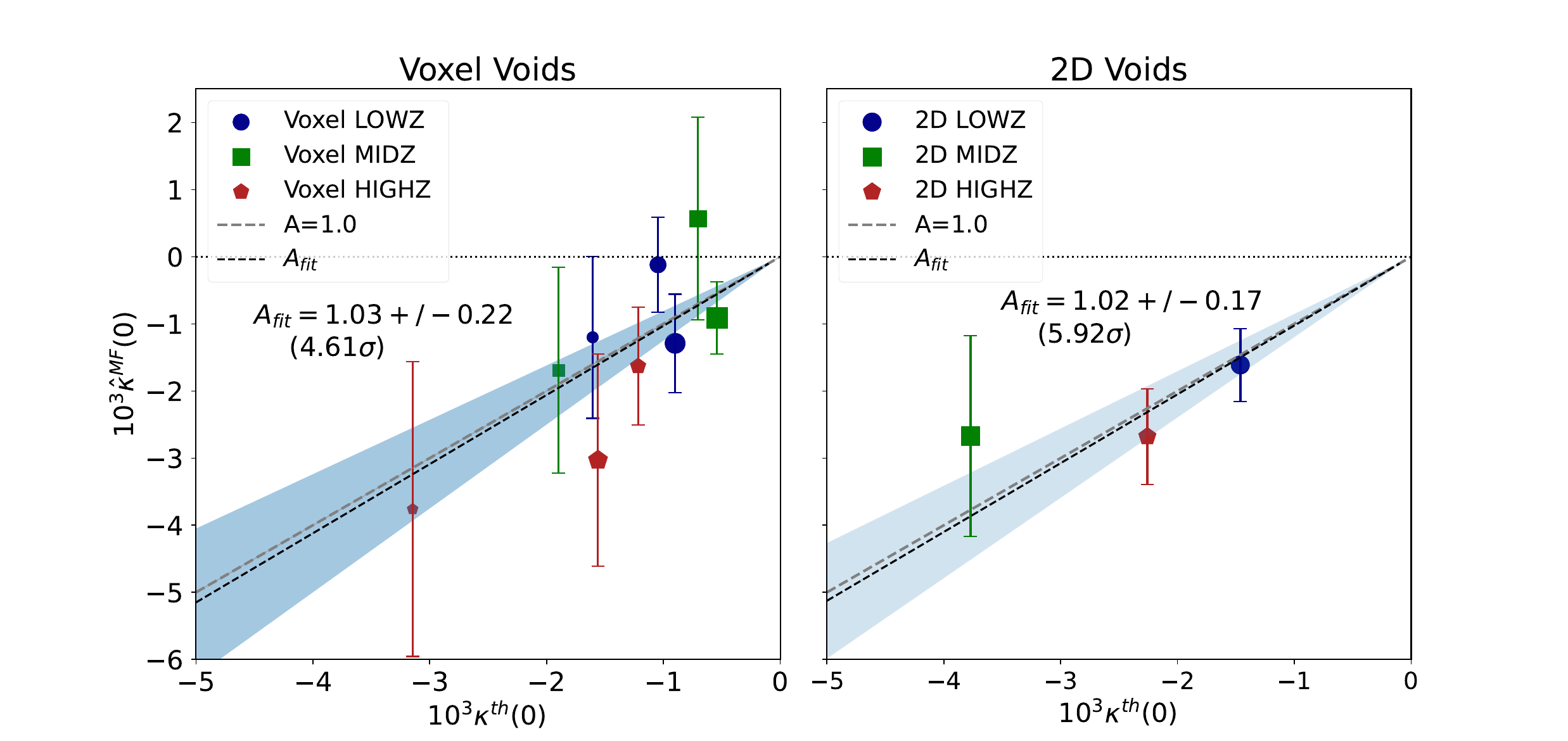}
    \caption{Comparison of \texttt{Voxel} and 2D Voids: This figure illustrates the correlation between the CMB lensing convergence ($\kappa$) derived from the Dark Energy Survey Year 3 (DESY3) and Planck data, and the simulated void lensing $\kappa$ from the MICE simulation at the stacked center pixels of voids. Each redshift bin (LOWZ, MIDZ, and HIGHZ) is represented by different markers. Within each redshift category, the bin values of $\lambda_v$ increase from the smallest to the largest. The plots include a reference line at $A_{k} =1.0$ and the best-fit line. Our analysis reveals a significant correlation between DESY3 and MICE data, with best-fit amplitudes of $A_{k} =1.03$ for \texttt{Voxel} voids and $A_{k} =1.02$ for 2D voids, corresponding to $4.61\sigma$ and $5.92\sigma$ detection levels for CMB lensing, respectively. These findings show a strong agreement with $\Lambda$CDM expectations.}
    \label{fig:final_result_void}
    \end{minipage}
\end{figure*}

\begin{figure*}
    \begin{minipage}{\textwidth}
    \centering
    \includegraphics[width=0.94\textwidth]{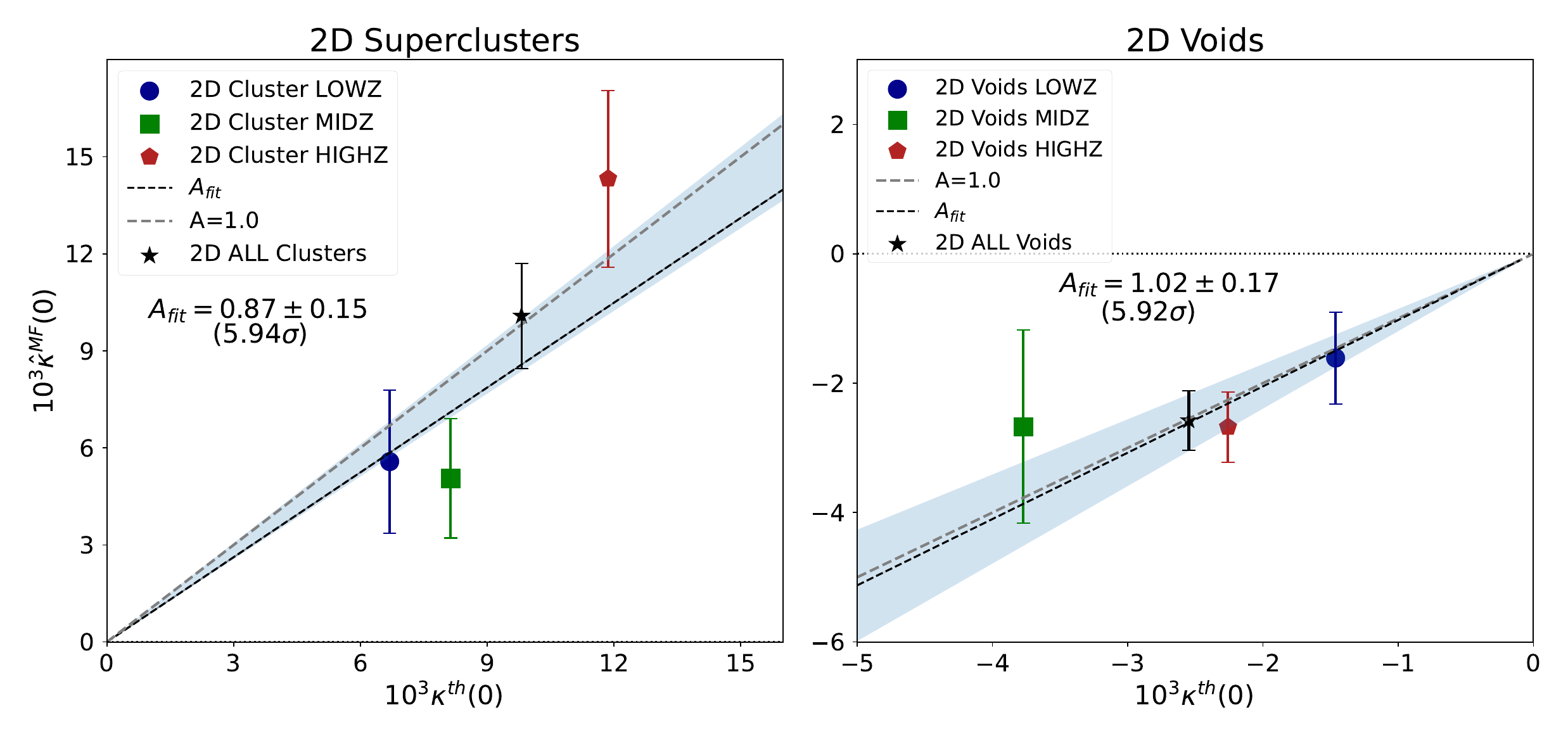}
    \caption{Comparative Analysis of 2D superclusters and 2D Voids: This figure presents stacked $\kappa_{0}$ values for each redshift bin of both 2D superclusters and 2D voids. Our joint fit utilizing different redshift categories yields a statistical significance of $5.94\sigma$ and an amplitude of $A_{k}=0.87 \pm 0.15$ for 2D superclusters, while for 2D voids we find a statistical significance of $5.92\sigma$ and an amplitude of $A_{k}=1.02 \pm 0.17$. The amplitude $A_{k}$ observed for 2D superclusters is slightly lower than that for 2D voids, but still falls within $1\sigma$ of $A_{k} =1.00$. These results are in line with \citep{Hang2021}, which also found that 2D superclusters exhibit a slightly lower $A_{k}$ value than their void counterparts. However, it's important to note that our measured $A_{k}$ values imply a good agreement with $A_{k}=1.00$, while \citep{Hang2021} demonstrate a marginally lower lensing amplitude for voids and superclusters.}
    \label{fig:final_result_cluster_void}
    \end{minipage}
\end{figure*}

The results of our study confirm that both \texttt{Voxel} and 2D void results are in good agreement with expectation, as seen in Figure~\ref{fig:final_result_void}. The measured $A_{k}$ values are $A_{Voxel} = 1.03 \pm 0.22$ and $A_{2D} = 1.02 \pm 0.17$. These represent a $4.61 \sigma$ detection of CMB lensing in the case of \texttt{Voxel} voids, and a $5.92 \sigma$ detection for 2D Voids. We attribute the marginally higher detection associated with 2D voids to their intrinsic elongation along the line of sight due to their projected nature in redshift shells.  We then apply the same redshift binning to compare the results of 2D superclusters as seen in Figure~\ref{fig:final_result_cluster_void}, finding \(A_{2D} = 0.87 \pm 0.15\), corresponding to a \(5.94 \sigma\) detection of CMB lensing.

\subsection{Comparison with K22 Voids}
\label{comparison_k22}
In order to understand the differences that the employed method and type of tracers can make, we compare our results with K22 \citep{kovacsy3cmb}. K22 uses the same dataset and simulation as in our study. The difference lies in our updated \textit{redMaGiC} algorithm to select the galaxies both in DES Y3 and the MICE simulation. 

In addition, we use the matched filtering method, and we do not re-scale when stacking the CMB cutouts around voids. However, to investigate if the employed method makes a significant difference in the results, we also use the same method as in the K22 study, rescaling the stacked CMB cutouts around void centers and filtering the CMB lensing maps with a Gaussian filter with a size Full Width at Half Maximum (FWHM) = 1$^{\circ}$. 

\begin{figure*}
    \begin{minipage}{\textwidth}
    \centering
    \includegraphics[width=0.75\textwidth]{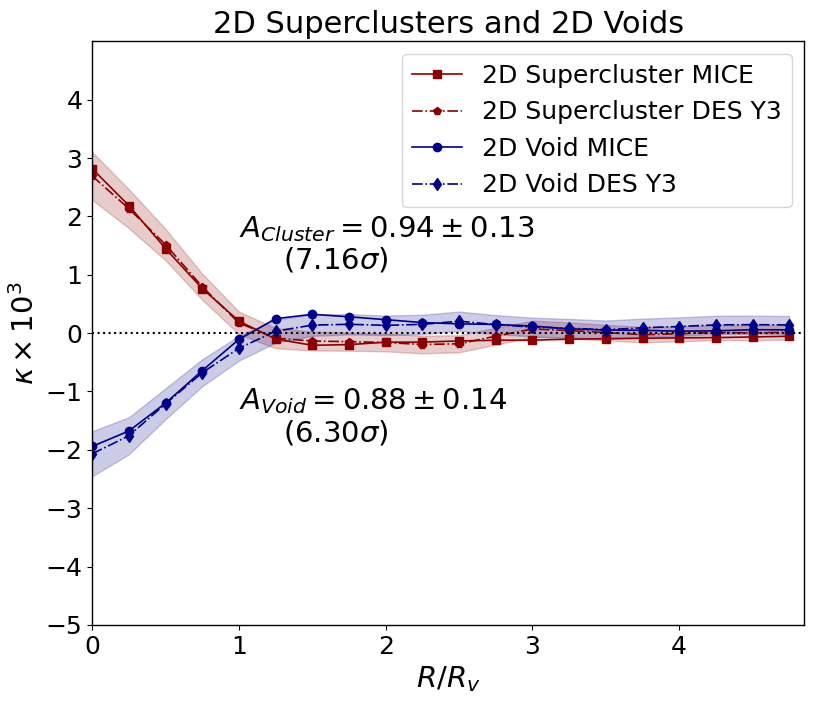}
    \caption{Stacked profile of 2D voids and 2D superclusters with the template fitting methodology from K22. The 2D superstructures with radius $R_{v} < 20\rvunits$ are not included in the analysis in accordance with K22.There is no further binning applied. The stacked images are obtained by rescaling the image up to  $5R_{v}$ centered on void positions. We then measured the radial kappa profile from the final stacked images. The shaded regions show the $1\sigma$ error bars from the covariance matrix as explained in Section\ref{comparison_k22} }
    \label{fig:rescaled_results}
    \end{minipage}
\end{figure*}

To determine the covariance matrix for this measurement, we generate 1000 random Planck maps in the same way as for our matched filtering approach, and repeat the Gaussian filtering and the re-scaled stacking of DES Y3 voids and superclusters, as carried out in K22. To estimate the lensing amplitude, we employ equation \ref{eq:chi2}, consistent with our matched filtering methodology. The templates and observed signals are shown in Figure~\ref{fig:rescaled_results}, and we find amplitudes \( A_{\kappa} = 0.88 \pm 0.14 \) (a \( 6.30\sigma \) detection of the lensing signal) for 2D voids, and \( A_{\kappa} = 0.94 \pm 0.13 \) (a \( 7.16\sigma \) detection) for 2D Clusters. Compared to the results of K22, who report values of $A_\kappa$ lower than the MICE expectation at around the $\sim2.3\sigma$ level, our results are perfectly consistent with $A_\kappa=1$. The statistical uncertainty in our results is slightly larger than that of K22, although very compatible.  We associate this difference between our result and that of K22 with the updated and improved \textit{redMaGiC} galaxy sample, which affected both the Y3 data and the MICE mocks. 

We further tested changing the cut used to define the superstructure sample, from $R_{v} < 20 \rvunits$ as used by K22 to $R_{v} < 15 \rvunits$ instead. This cut naturally substantially increases the number of 2D voids and 2D superclusters in the final sample, and leads to a reduction in the measurement uncertainties of $\sim15\%$ in each case, while leaving the $A_\kappa$ central values essentially unchanged. This suggests that the K22 size cut is not optimal, but we leave a fuller investigation of optimisation of the signal-to-noise to future work.

\begin{table*}
\centering
\caption{Comparison of 2D Voids and 2D Superclusters in MICE and DES Y3. Total numbers without any cut are as follows: MICE 2D Voids: 12610, MICE 2D Superclusters: 13167, DES Y3 2D Voids: 10904, DES Y3 2D Superclusters: 10592.}
\begin{tabular}{ccccccc}
\hline
Category & Redshift & MICE (>20Mpc/h) & DES Y3 (>20Mpc/h) \\
\hline
2D Voids & Low & 1201 & 921 \\
& Mid & 1840 & 1512 \\
& High & 3254 & 2715 \\
& Total (>20Mpc/h) & 6295 & 5148 \\
\hline
2D Superclusters & Low & 1043 & 809 \\
& Mid & 1556 & 1338 \\
& High & 2885 & 2285 \\
& Total (>20Mpc/h) & 5484 & 4432 \\
\hline
\end{tabular}
\label{tab:comparison_2d}
\end{table*}

%%%%%%%%%%%%%%%%%%%%%%%%%%%%%%%%%%%%%%%%%%%%
%%%%%%%%%%%%%%%%%%%%%%%%%%%%%%%%%%%%%%%%%%%%
%%%%%%%%%%%%%%%%%%%%%%%%%%%%%%%%%%%%%%%%%%%%
%%%%%%%%%%%%%%%%%%%%%%%%%%%%%%%%%%%%%%%%%%%%

\section{Discussion and Conclusion} \label{Discussion and Conclusion}

In this study, we conducted an in-depth examination of superstructures and their interrelation with the CMB lensing, zeroing in on a pivotal instrument for the detection of the CMB lensing effect: the matched filter technique.

For the first time in the literature, we applied the \texttt{Voxel} void-finding algorithm to photo-z galaxies and gauged their CMB lensing footprints.

The CMB lensing footprints of cosmic voids, sourced from both the DES Y3 dataset and the MICE simulation for \texttt{Voxel} and 2D voids, demonstrated a robust agreement. This revelation provides compelling insights into the intricacies of delineating superstructures. It highlights the significance of not only the methodologies employed but also the noise properties associated with CMB lensing map and the choice of galaxies utilized in their definition.

Our results indicate a slight divergence from the weaker signal reported for DES Y3 voids in the study by \citep{kovacsy3cmb}, but align well with the findings for DES Y1 voids by \citep{vielzeufy1}. The disparity with the previous DES Y3 results is primarily attributed to the enhancements made in the \textit{redMaGiC} galaxy selection algorithm employed in our investigation. It became evident that the selection of tracer galaxies used to define superstructures wields a considerable influence on the CMB lensing outcomes of these, independent of the methodology employed in identifying the superstructures.

Our research emphasized significant observations of lensing stemming from \texttt{Voxel} voids, 2D voids, and 2D clusters, each uniformly demonstrating an amplitude consistent with $A_{k}=1.00$. Such uniformity across a variety of void types and structures not only reinforces the robustness of our findings but also aligns seamlessly with the predictions of the $\Lambda$CDM cosmological model, agreeing well with the results of previous studies \citep{raghunathan19,vielzeufy1,cai17,Hang2021,gisela}.

Furthermore, we identified a notable correspondence between 2D and 3D (\texttt{Voxel}) voids in our study, where the 2D voids exhibited marginally lower uncertainties and enhanced levels of CMB lensing detection. This observation indicates the superior precision of 2D void analysis in capturing lensing signals. Such precision is presumably owed to the elongated structure of 2D voids along the line of sight, which are defined on projected tomographic redshift slices. The 2D algorithm finds structures elongated along the line of sight because they result in a big impact on the projected density field but structures aligned perpendicular do not. This structural characteristic possibly contributes to the increased sensitivity in detecting lensing phenomena which is the integral effect along the line of sight, further validating the efficacy of employing 2D void analysis in studies of this nature.

Moreover, we demonstrated that \texttt{Voxel} voids can be effectively utilized for analyses involving the cross-correlation of lensing effects on the CMB. This study is the first to measure the CMB lensing imprints of \texttt{Voxel} voids. Our results provide evidence that these voids exist in regions of the Universe characterized by a true deficit in matter density as shown by their CMB lensing imprints. Encouragingly, our results align closely with those derived from 2D voids, a void finder methodology employed extensively in previous studies. Furthermore, as demonstrated by \citep{radinovicvoxel}, \texttt{Voxel} voids serve as a viable choice for void-galaxy correlation analyses. Unlike 2D voids, whose tomographically projected nature precludes this type of analysis, \texttt{Voxel} voids do not suffer from this limitation. This key advantage makes \texttt{Voxel} voids particularly suited to forthcoming large-scale surveys such as the Dark Energy Spectroscopic Instrument (DESI) \citep{desia,desib} and the Euclid mission \citep{euclid1}.

In our comparison of the matched filter and Gaussian filtering methods, we observed consistent results for $A_{k} =1.00$ across 2D voids and clusters, as detailed in \citep{kovacsy3cmb,vielzeufy1}. Our study demonstrated that the matched filter method can provide more precise $A_{k}$ estimates, suggesting its potential for future research. In particular, it can be interesting to utilize the matched filtering method on the same dataset used in \citep{kovacs22isw} to examine the detected ISW signals. The authors of the cited study reported an excess ISW signal without using matched filtering, which presents an intriguing opportunity for future analysis and verification of their claim.

In the near future, the final processed data of the full six years of DES observations will be available for such analyses. The DES Y6 dataset covers the same footprint as the Y3 data used here, but is deeper and thus has a factor of $\sim2.3\times$ higher number density of galaxy tracers. Given the fixed footprint, we expect only a small change in the number of 2D voids in Y6, as these are less sensitive to the galaxy number density. In contrast, for \texttt{Voxel} voids, the number of voids is expected to change roughly proportional to the change in the tracer number density, so we expect a factor of $\sim2.3\times$ increase in the number of voids for Y6. However, this does not translate to a simple  $1/\sqrt{N_\mathrm{voids}}$  reduction in the statistical uncertainty, because the nature of individual voids found also changes as the galaxy number density increases, with smaller voids being resolved. This means that the expected signal strength also changes. A full accounting of the net effect on the expected SNR requires dedicated studies using Y6 mocks.

This work provides the pivotal groundwork for upcoming studies from the Vera Rubin Observatory and the Euclid Survey, which aim to further investigate the Integrated Sachs-Wolfe (ISW) effect and CMB lensing due to cosmic superstructures. Our research not only contributes to the evolving understanding of lensing phenomena in the Universe but also lays a strong foundation for future studies. By applying the insights discussed in our analysis, future large-scale structure surveys can refine their strategies for exploring the universe's large-scale structure, potentially leading to enhanced accuracy and precision in cosmological constraints derived from these measurements.

%%%%%%%%%%%%%%%%%%%%%%%%%%%%%%%%%%%%%%%%%%%%
%%%%%%%%%%%%%%%%%%%%%%%%%%%%%%%%%%%%%%%%%%%%
%%%%%%%%%%%%%%%%%%%%%%%%%%%%%%%%%%%%%%%%%%%%
%%%%%%%%%%%%%%%%%%%%%%%%%%%%%%%%%%%%%%%%%%%%%%%%%%%%%%%

%%%%%%%%%%%%%%%%%%%%%%%%%%%%%%%%%%%%%%%%%%%%%%%%%%
%%%%%%%%%%%%%%%%%%%%%%%%%%%%%%%%%%%%%%%%%%%%%%%%%%
%%%%%%%%%%%%%%%%%%%%%%%%%%%%%%%%%%%%%%%%%%%%%%%%%%
%%%%%%%%%%%%%%%%%%%%%%%%%%%%%%%%%%%%%%%%%%%%%%%%%%
\section*{Acknowledgements}

\textbf{Author contributions:} We would like to acknowledge everyone who made this work possible. UD and SN made significant contributions to project development, including paper writing and figures. They also were involved in data analysis and methods validation. IF was instrumental in the construction and validation of the MICE redMaGiC catalog, ensuring the accuracy and reliability of this critical dataset. PF took charge of constructing and validating the MICE CMB lensing map. CTD and SP conducted an internal review of the paper within the DES collaboration, offering suggestions that significantly improved its quality. AK was pivotal in idea generation and providing useful suggestions that helped shape the project's direction. SN and RM provided advice, guiding the project through their expertise in the field. MA, KB, ADW, RAG, WGH, APi, ESR, ESh and BY were responsible for the construction and validation of the DES redMagic catalog, a cornerstone of our analysis. The remaining authors have made contributions to this paper that include, but are not limited to, the construction of DECam and other aspects of data collection; data processing and calibration; development of broadly used methods, codes, and simulations; operation of the pipelines and validation tests; and advancement of the science analysis.

UD acknowledges the support by PREBIST project and has received funding from the European Union’s Horizon 2020 research and innovation programme under the Marie Skłodowska-Curie grant agreement No.754558. 
SN acknowledges support from an STFC Ernest Rutherford Fellowship, grant reference ST/T005009/2. AK received funding from the European Union’s Horizon Europe research and innovation programme under the Marie Skłodowska-Curie grant agreement number 101130774, from the Hungarian Ministry of Innovation and Technology NRDI Office grant OTKA NN147550, and from a \emph{Lend\"ulet} excellence grant by the Hungarian Academy of Sciences (MTA).

Funding for the DES Projects has been provided by the U.S. Department of Energy, the U.S. National Science Foundation, the Ministry of Science and Education of Spain, the Science and Technology Facilities Council of the United Kingdom, the Higher Education Funding Council for England, the National Center for Supercomputing Applications at the University of Illinois at Urbana-Champaign, the Kavli Institute of Cosmological Physics at the University of Chicago, the Center for Cosmology and Astro-Particle Physics at the Ohio State University, the Mitchell Institute for Fundamental Physics and Astronomy at Texas AM University, Financiadora de Estudos e Projetos, Fundação Carlos Chagas Filho de Amparo à Pesquisa do Estado do Rio de Janeiro, Conselho Nacional de Desenvolvimento Científico e Tecnológico and the Ministério da Ciência, Tecnologia e Inovação, the Deutsche Forschungsgemeinschaft and the Collaborating Institutions in the Dark Energy Survey.  The Collaborating Institutions are Argonne National Laboratory, the University of California at Santa Cruz, the University of Cambridge, Centro de Investigaciones Energéticas, Medioambientales y Tecnológicas-Madrid, the University of Chicago, University College London, the DES-Brazil Consortium, the University of Edinburgh, the Eidgenössische Technische Hochschule (ETH) Zürich, Fermi National Accelerator Laboratory, the University of Illinois at Urbana-Champaign, the Institut de Ciències de l’Espai (IEEC/CSIC), the Institut de Física d’Altes Energies, Lawrence Berkeley National Laboratory, the Ludwig-Maximilians Universität München and the associated Excellence Cluster Universe, the University of Michigan, NSF’s NOIRLab, the University of Nottingham, The Ohio State University, the University of Pennsylvania, the University of Portsmouth, SLAC National Accelerator Laboratory, Stanford University, the University of Sussex, Texas AM University, and the OzDES Membership Consortium. Based in part on observations at Cerro Tololo InterAmerican Observatory at NSF’s NOIRLab (NOIRLab Prop. ID 2012B-0001; PI: J. Frieman), which is managed by the Association of Universities for Research in Astronomy (AURA) under a cooperative agreement with the National Science Foundation.

The DES data management system is supported by the National Science Foundation under Grant Numbers AST-1138766 and AST-1536171. The DES participants from Spanish institutions are partially supported by MICINN under grants ESP2017-89838, PGC2018-094773, PGC2018-102021, SEV-2016-0588, SEV-2016-0597, and MDM-2015-0509, some of which include ERDF funds from the European Union. IFAE is partially funded by the CERCA program of the Generalitat de Catalunya. Research leading to these results has received funding from the European Research Council under the European Union’s Seventh Framework Program (FP7/2007-2013) including ERC grant agreements 240672, 291329,and 306478. We acknowledge support from the Brazilian Instituto Nacional de Ciência e Tecnologia (INCT) do e-Universo (CNPq grant 465376/2014-2).

The data production, processing and analysis tools for this paper have been developed, implemented and operated in collaboration with the Port d’Informació Científica (PIC) data center. PIC is maintained through a collaboration agreement between the Institut de Física d’Altes Energies (IFAE) and the Centro de Investigaciones Energéticas, Medioambientales y Tecnológicas (CIEMAT). UD specifically thanks Carles Acosta and Christian Neissner for their invaluable assistance with the PIC services, and warmly acknowledges the enlightening discussions with Qianjun Hang that have contributed to this research.

\section*{Data Availability}
This research uses data from the Dark Energy Survey Y3 dataset (publicly available at \url{https://des.ncsa.illinois.edu/releases/y3a2}) and from the {\it Planck} mission (publicly available at \url{https://pla.esac.esa.int/pla/#home}). Additional data products specific to this paper are available upon request from the corresponding author.

%%%%%%%%%%%%%%%%%%%% REFERENCES %%%%%%%%%%%%%%%%%%
% The best way to enter references is to use BibTeX:
\bibliographystyle{mnras}
\bibliography{demirbozan} % if your bibtex file is called 

\begin{thebibliography}{}
\makeatletter
\relax
\def\mn@urlcharsother{\let\do\@makeother \do\$\do\&\do\#\do\^\do\_\do\%\do\~}
\def\mn@doi{\begingroup\mn@urlcharsother \@ifnextchar [ {\mn@doi@} {\mn@doi@[]}}
\def\mn@doi@[#1]#2{\def\@tempa{#1}\ifx\@tempa\@empty \href {http://dx.doi.org/#2} {doi:#2}\else \href {http://dx.doi.org/#2} {#1}\fi \endgroup}
\def\mn@eprint#1#2{\mn@eprint@#1:#2::\@nil}
\def\mn@eprint@arXiv#1{\href {http://arxiv.org/abs/#1} {{\tt arXiv:#1}}}
\def\mn@eprint@dblp#1{\href {http://dblp.uni-trier.de/rec/bibtex/#1.xml} {dblp:#1}}
\def\mn@eprint@#1:#2:#3:#4\@nil{\def\@tempa {#1}\def\@tempb {#2}\def\@tempc {#3}\ifx \@tempc \@empty \let \@tempc \@tempb \let \@tempb \@tempa \fi \ifx \@tempb \@empty \def\@tempb {arXiv}\fi \@ifundefined {mn@eprint@\@tempb}{\@tempb:\@tempc}{\expandafter \expandafter \csname mn@eprint@\@tempb\endcsname \expandafter{\@tempc}}}

\bibitem[\protect\citeauthoryear{Abdalla et~al.,}{Abdalla et~al.}{2022}]{snowmasstension}
Abdalla E.,  et~al., 2022, \mn@doi [Journal of High Energy Astrophysics] {10.1016/j.jheap.2022.04.002}, 34, 49

\bibitem[\protect\citeauthoryear{{Ade} et~al.,}{{Ade} et~al.}{2019}]{SO19}
{Ade} P.,  et~al., 2019, \mn@doi [\jcap] {10.1088/1475-7516/2019/02/056}, \href {https://ui.adsabs.harvard.edu/abs/2019JCAP...02..056A} {2019, 056}

\bibitem[\protect\citeauthoryear{{Baxter} et~al.,}{{Baxter} et~al.}{2015}]{baxter15}
{Baxter} E.~J.,  et~al., 2015, \mn@doi [\apj] {10.1088/0004-637X/806/2/247}, \href {http://adsabs.harvard.edu/abs/2015ApJ...806..247B} {806, 247}

\bibitem[\protect\citeauthoryear{{Cai}, {Neyrinck}, {Mao}, {Peacock}, {Szapudi}  \& {Berlind}}{{Cai} et~al.}{2017}]{cai17}
{Cai} Y.-C.,  {Neyrinck} M.,  {Mao} Q.,  {Peacock} J.~A.,  {Szapudi} I.,   {Berlind} A.~A.,  2017, \mn@doi [\mnras] {10.1093/mnras/stw3299}, \href {https://ui.adsabs.harvard.edu/abs/2017MNRAS.466.3364C} {466, 3364}

\bibitem[\protect\citeauthoryear{Camacho-Ciurana, Lee, Arsenov, Kovács, Szapudi  \& Csabai}{Camacho-Ciurana et~al.}{2023}]{gisela}
Camacho-Ciurana G.,  Lee P.,  Arsenov N.,  Kovács A.,  Szapudi I.,   Csabai I.,  2023, The CMB lensing imprint of cosmic voids detected in the WISE-Pan-STARRS luminous red galaxy catalog (\mn@eprint {arXiv} {2312.08483})

\bibitem[\protect\citeauthoryear{{Clampitt} \& {Jain}}{{Clampitt} \& {Jain}}{2015}]{clampitt15}
{Clampitt} J.,  {Jain} B.,  2015, \mn@doi [\mnras] {10.1093/mnras/stv2215}, \href {https://ui.adsabs.harvard.edu/abs/2015MNRAS.454.3357C} {454, 3357}

\bibitem[\protect\citeauthoryear{{Crocce}, {Castander}, {Gazta{\~n}aga}, {Fosalba}  \& {Carretero}}{{Crocce} et~al.}{2015}]{crocce2015}
{Crocce} M.,  {Castander} F.~J.,  {Gazta{\~n}aga} E.,  {Fosalba} P.,   {Carretero} J.,  2015, \mn@doi [\mnras] {10.1093/mnras/stv1708}, \href {http://adsabs.harvard.edu/abs/2015MNRAS.453.1513C} {453, 1513}

\bibitem[\protect\citeauthoryear{{DESI Collaboration} et~al.,}{{DESI Collaboration} et~al.}{2016a}]{desia}
{DESI Collaboration} et~al., 2016a, \mn@doi [arXiv e-prints] {10.48550/arXiv.1611.00036}, \href {https://ui.adsabs.harvard.edu/abs/2016arXiv161100036D} {p. arXiv:1611.00036}

\bibitem[\protect\citeauthoryear{{DESI Collaboration} et~al.,}{{DESI Collaboration} et~al.}{2016b}]{desib}
{DESI Collaboration} et~al., 2016b, \mn@doi [arXiv e-prints] {10.48550/arXiv.1611.00037}, \href {https://ui.adsabs.harvard.edu/abs/2016arXiv161100037D} {p. arXiv:1611.00037}

\bibitem[\protect\citeauthoryear{{Dark Energy Survey Collaboration}}{{Dark Energy Survey Collaboration}}{2016}]{morethanDE2016}
{Dark Energy Survey Collaboration} 2016, \mn@doi [\mnras] {10.1093/mnras/stw641}, \href {http://adsabs.harvard.edu/abs/2016MNRAS.460.1270D} {460, 1270}

\bibitem[\protect\citeauthoryear{{Dunkley} et~al.,}{{Dunkley} et~al.}{2009}]{WMAPparam2009}
{Dunkley} J.,  et~al., 2009, \mn@doi [\apjs] {10.1088/0067-0049/180/2/306}, \href {http://adsabs.harvard.edu/abs/2009ApJS..180..306D} {180, 306}

\bibitem[\protect\citeauthoryear{{Fang} et~al.,}{{Fang} et~al.}{2019}]{Fang2019}
{Fang} Y.,  et~al., 2019, \mn@doi [\mnras] {10.1093/mnras/stz2805}, \href {https://ui.adsabs.harvard.edu/abs/2019MNRAS.tmp.2404F} {p.~2404}

\bibitem[\protect\citeauthoryear{{Flaugher}, {Diehl}, {Honscheid}  \& {The DES Collaboration}}{{Flaugher} et~al.}{2015}]{DECam}
{Flaugher} B.,  {Diehl} H.~T.,  {Honscheid} K.,   {The DES Collaboration} 2015, \mn@doi [\aj] {10.1088/0004-6256/150/5/150}, \href {http://adsabs.harvard.edu/abs/2015AJ....150..150F} {150, 150}

\bibitem[\protect\citeauthoryear{{Fosalba}, {Gazta{\~n}aga}, {Castander}  \& {Manera}}{{Fosalba} et~al.}{2008}]{fosalba2008onion}
{Fosalba} P.,  {Gazta{\~n}aga} E.,  {Castander} F.~J.,   {Manera} M.,  2008, \mn@doi [\mnras] {10.1111/j.1365-2966.2008.13910.x}, \href {http://adsabs.harvard.edu/abs/2008MNRAS.391..435F} {391, 435}

\bibitem[\protect\citeauthoryear{{Fosalba}, {Gazta{\~n}aga}, {Castander}  \& {Crocce}}{{Fosalba} et~al.}{2015a}]{fosalba2015}
{Fosalba} P.,  {Gazta{\~n}aga} E.,  {Castander} F.~J.,   {Crocce} M.,  2015a, \mn@doi [\mnras] {10.1093/mnras/stu2464}, \href {http://adsabs.harvard.edu/abs/2015MNRAS.447.1319F} {447, 1319}

\bibitem[\protect\citeauthoryear{{Fosalba}, {Crocce}, {Gazta{\~n}aga}  \& {Castander}}{{Fosalba} et~al.}{2015b}]{fosalba2015a}
{Fosalba} P.,  {Crocce} M.,  {Gazta{\~n}aga} E.,   {Castander} F.~J.,  2015b, \mn@doi [\mnras] {10.1093/mnras/stv138}, \href {http://adsabs.harvard.edu/abs/2015MNRAS.448.2987F} {448, 2987}

\bibitem[\protect\citeauthoryear{{G{\'o}rski}, {Hivon}, {Banday}, {Wand elt}, {Hansen}, {Reinecke}  \& {Bartelmann}}{{G{\'o}rski} et~al.}{2005}]{gorski05}
{G{\'o}rski} K.~M.,  {Hivon} E.,  {Banday} A.~J.,  {Wand elt} B.~D.,  {Hansen} F.~K.,  {Reinecke} M.,   {Bartelmann} M.,  2005, \mn@doi [\apj] {10.1086/427976}, \href {https://ui.adsabs.harvard.edu/abs/2005ApJ...622..759G} {622, 759}

\bibitem[\protect\citeauthoryear{Granett, Neyrinck  \& Szapudi}{Granett et~al.}{2008}]{Granett:2008}
Granett B.~R.,  Neyrinck M.~C.,   Szapudi I.,  2008, \mn@doi [\apj] {10.1086/591670}, 683, L99

\bibitem[\protect\citeauthoryear{{Gruen}, {Friedrich}, {Amara}, {Bacon}, {Bonnett}  \& et al.}{{Gruen} et~al.}{2016}]{Gruen2016}
{Gruen} D.,  {Friedrich} O.,  {Amara} A.,  {Bacon} D.,  {Bonnett} C.,   et al. 2016, \mn@doi [\mnras] {10.1093/mnras/stv2506}, \href {http://adsabs.harvard.edu/abs/2016MNRAS.455.3367G} {455, 3367}

\bibitem[\protect\citeauthoryear{{Hamaus}, {Pisani}, {Sutter}, {Lavaux}, {Escoffier}, {Wandelt}  \& {Weller}}{{Hamaus} et~al.}{2016}]{Hamaus:2016}
{Hamaus} N.,  {Pisani} A.,  {Sutter} P.~M.,  {Lavaux} G.,  {Escoffier} S.,  {Wandelt} B.~D.,   {Weller} J.,  2016, \mn@doi [Physical Review Letters] {10.1103/PhysRevLett.117.091302}, \href {http://adsabs.harvard.edu/abs/2016PhRvL.117i1302H} {117, 091302}

\bibitem[\protect\citeauthoryear{{Hamaus}, {Cousinou}, {Pisani}, {Aubert}, {Escoffier}  \& {Weller}}{{Hamaus} et~al.}{2017}]{Hamaus:2017a}
{Hamaus} N.,  {Cousinou} M.-C.,  {Pisani} A.,  {Aubert} M.,  {Escoffier} S.,   {Weller} J.,  2017, \mn@doi [\jcap] {10.1088/1475-7516/2017/07/014}, \href {http://adsabs.harvard.edu/abs/2017JCAP...07..014H} {7, 014}

\bibitem[\protect\citeauthoryear{Hang, Alam, Cai  \& Peacock}{Hang et~al.}{2021}]{Hang2021}
Hang Q.,  Alam S.,  Cai Y.-C.,   Peacock J.~A.,  2021, \mn@doi [\mnras] {10.1093/mnras/stab2184}, 507, 510

\bibitem[\protect\citeauthoryear{{Hartlap}, {Simon}  \& {Schneider}}{{Hartlap} et~al.}{2007}]{hartlap2007}
{Hartlap} J.,  {Simon} P.,   {Schneider} P.,  2007, \mn@doi [\aap] {10.1051/0004-6361:20066170}, \href {http://adsabs.harvard.edu/abs/2007A%26A...464..399H} {464, 399}

\bibitem[\protect\citeauthoryear{{He}, {Alam}, {Ferraro}, {Chen}  \& {Ho}}{{He} et~al.}{2018}]{he18}
{He} S.,  {Alam} S.,  {Ferraro} S.,  {Chen} Y.-C.,   {Ho} S.,  2018, \mn@doi [Nature Astronomy] {10.1038/s41550-018-0426-z}, \href {https://ui.adsabs.harvard.edu/abs/2018NatAs...2..401H} {2, 401}

\bibitem[\protect\citeauthoryear{{Heymans} et~al.,}{{Heymans} et~al.}{2021}]{heymanskids}
{Heymans} C.,  et~al., 2021, \mn@doi [\aap] {10.1051/0004-6361/202039063}, \href {https://ui.adsabs.harvard.edu/abs/2021A&A...646A.140H} {646, A140}

\bibitem[\protect\citeauthoryear{{Hotchkiss}, {Nadathur}, {Gottl{\"o}ber}, {Iliev}, {Knebe}, {Watson}  \& {Yepes}}{{Hotchkiss} et~al.}{2015}]{Hotchkiss2015}
{Hotchkiss} S.,  {Nadathur} S.,  {Gottl{\"o}ber} S.,  {Iliev} I.~T.,  {Knebe} A.,  {Watson} W.~A.,   {Yepes} G.,  2015, \mn@doi [\mnras] {10.1093/mnras/stu2072}, \href {http://adsabs.harvard.edu/abs/2015MNRAS.446.1321H} {446, 1321}

\bibitem[\protect\citeauthoryear{{Hu} \& {Okamoto}}{{Hu} \& {Okamoto}}{2002}]{hu02}
{Hu} W.,  {Okamoto} T.,  2002, \mn@doi [\apj] {10.1086/341110}, \href {https://ui.adsabs.harvard.edu/abs/2002ApJ...574..566H} {574, 566}

\bibitem[\protect\citeauthoryear{{Khoury} \& {Weltman}}{{Khoury} \& {Weltman}}{2004}]{khoury2004}
{Khoury} J.,  {Weltman} A.,  2004, \mn@doi [\prd] {10.1103/PhysRevD.69.044026}, \href {http://adsabs.harvard.edu/abs/2004PhRvD..69d4026K} {69, 044026}

\bibitem[\protect\citeauthoryear{{Kitaura} et~al.,}{{Kitaura} et~al.}{2016}]{Kitaura2016}
{Kitaura} F.-S.,  et~al., 2016, \mn@doi [\prl] {10.1103/PhysRevLett.116.171301}, \href {https://ui.adsabs.harvard.edu/abs/2016PhRvL.116q1301K} {116, 171301}

\bibitem[\protect\citeauthoryear{{Kov{\'a}cs}}{{Kov{\'a}cs}}{2018}]{kovacs2017}
{Kov{\'a}cs} A.,  2018, \mn@doi [\mnras] {10.1093/mnras/stx3213}, \href {http://adsabs.harvard.edu/abs/2018MNRAS.475.1777K} {475, 1777}

\bibitem[\protect\citeauthoryear{{Kov{\'a}cs} et~al.,}{{Kov{\'a}cs} et~al.}{2019a}]{Kovacs2019more}
{Kov{\'a}cs} A.,  et~al., 2019a, \mn@doi [\mnras] {10.1093/mnras/stz341}, \href {https://ui.adsabs.harvard.edu/abs/2019MNRAS.484.5267K} {484, 5267}

\bibitem[\protect\citeauthoryear{{Kov{\'a}cs}, {S{\'a}nchez}, {Garc{\'{\i}}a-Bellido}  \& the DES~collaboration}{{Kov{\'a}cs} et~al.}{2019b}]{Kovacs2019}
{Kov{\'a}cs} A.,  {S{\'a}nchez} C.,  {Garc{\'{\i}}a-Bellido} J.,   the DES~collaboration 2019b, \mn@doi [\mnras] {10.1093/mnras/stz341}, \href {https://ui.adsabs.harvard.edu/abs/2019MNRAS.484.5267K} {484, 5267}

\bibitem[\protect\citeauthoryear{Kovacs et~al.,}{Kovacs et~al.}{2021}]{kovacs2021eridanus}
Kovacs A.,  et~al., 2021, \mn@doi [\mnras] {10.1093/mnras/stab3309}, 510, 216

\bibitem[\protect\citeauthoryear{Kov{\'a}cs, Beck, Smith, R{\'{a}}cz, Csabai  \& Szapudi}{Kov{\'a}cs et~al.}{2022a}]{kovacs22isw}
Kov{\'a}cs A.,  Beck R.,  Smith A.,  R{\'{a}}cz G.,  Csabai I.,   Szapudi I.,  2022a, \mn@doi [\mnras] {10.1093/mnras/stac903}, 513, 15

\bibitem[\protect\citeauthoryear{{Kov{\'a}cs} et~al.,}{{Kov{\'a}cs} et~al.}{2022b}]{kovacsy3cmb}
{Kov{\'a}cs} A.,  et~al., 2022b, \mn@doi [\mnras] {10.1093/mnras/stac2011}, \href {https://ui.adsabs.harvard.edu/abs/2022MNRAS.515.4417K} {515, 4417}

\bibitem[\protect\citeauthoryear{{Kreisch}, {Pisani}, {Carbone}, {Liu}, {Hawken}, {Massara}, {Spergel}  \& {Wandelt}}{{Kreisch} et~al.}{2019}]{kreisch19}
{Kreisch} C.~D.,  {Pisani} A.,  {Carbone} C.,  {Liu} J.,  {Hawken} A.~J.,  {Massara} E.,  {Spergel} D.~N.,   {Wandelt} B.~D.,  2019, \mn@doi [\mnras] {10.1093/mnras/stz1944}, \href {https://ui.adsabs.harvard.edu/abs/2019MNRAS.488.4413K} {488, 4413}

\bibitem[\protect\citeauthoryear{{Laureijs} et~al.,}{{Laureijs} et~al.}{2011}]{euclid1}
{Laureijs} R.,  et~al., 2011, \mn@doi [arXiv e-prints] {10.48550/arXiv.1110.3193}, \href {https://ui.adsabs.harvard.edu/abs/2011arXiv1110.3193L} {p. arXiv:1110.3193}

\bibitem[\protect\citeauthoryear{Lavaux \& Wandelt}{Lavaux \& Wandelt}{2012}]{Lavaux2012}
Lavaux G.,  Wandelt B.~D.,  2012, \mn@doi [\apj] {10.1088/0004-637x/754/2/109}, 754, 109

\bibitem[\protect\citeauthoryear{Lesgourgues \& Pastor}{Lesgourgues \& Pastor}{2006}]{lesg2006}
Lesgourgues J.,  Pastor 2006, \mn@doi [Physics Reports] {10.1016/j.physrep.2006.04.001}, 429, 307

\bibitem[\protect\citeauthoryear{Lesgourgues, Mangano, Miele  \& Pastor}{Lesgourgues et~al.}{2013}]{lesg2013}
Lesgourgues J.,  Mangano G.,  Miele G.,   Pastor S.,  2013, {Neutrino Cosmology}.
Cambridge University Press

\bibitem[\protect\citeauthoryear{{Li}, {Zhao}  \& {Koyama}}{{Li} et~al.}{2012}]{li2012}
{Li} B.,  {Zhao} G.-B.,   {Koyama} K.,  2012, \mn@doi [\mnras] {10.1111/j.1365-2966.2012.20573.x}, \href {http://adsabs.harvard.edu/abs/2012MNRAS.421.3481L} {421, 3481}

\bibitem[\protect\citeauthoryear{{Madhavacheril} et~al.,}{{Madhavacheril} et~al.}{2015}]{madhavacheril15}
{Madhavacheril} M.,  et~al., 2015, \mn@doi [Physical Review Letters] {10.1103/PhysRevLett.114.151302}, \href {http://adsabs.harvard.edu/abs/2015PhRvL.114o1302M} {114, 151302}

\bibitem[\protect\citeauthoryear{{Martino} \& {Sheth}}{{Martino} \& {Sheth}}{2009}]{Martinosheth2009}
{Martino} M.~C.,  {Sheth} R.~K.,  2009, preprint, \href {http://adsabs.harvard.edu/abs/2009arXiv0911.1829M} {} (\mn@eprint {arXiv} {0911.1829})

\bibitem[\protect\citeauthoryear{{Massara}, {Villaescusa-Navarro}, {Viel}  \& {Sutter}}{{Massara} et~al.}{2015}]{massara15}
{Massara} E.,  {Villaescusa-Navarro} F.,  {Viel} M.,   {Sutter} P.~M.,  2015, \mn@doi [\jcap] {10.1088/1475-7516/2015/11/018}, \href {https://ui.adsabs.harvard.edu/abs/2015JCAP...11..018M} {2015, 018}

\bibitem[\protect\citeauthoryear{{McEwen}, {Hobson}  \& {Lasenby}}{{McEwen} et~al.}{2008}]{McEwen:2008}
{McEwen} J.~D.,  {Hobson} M.~P.,   {Lasenby} A.~N.,  2008, \mn@doi [IEEE Transactions on Signal Processing] {10.1109/TSP.2008.923198}, \href {http://adsabs.harvard.edu/abs/2008ITSP...56.3813M} {56, 3813}

\bibitem[\protect\citeauthoryear{{Melchior}, {Sutter}, {Sheldon}, {Krause}  \& {Wandelt}}{{Melchior} et~al.}{2014}]{melchior14}
{Melchior} P.,  {Sutter} P.~M.,  {Sheldon} E.~S.,  {Krause} E.,   {Wandelt} B.~D.,  2014, \mn@doi [\mnras] {10.1093/mnras/stu456}, \href {https://ui.adsabs.harvard.edu/abs/2014MNRAS.440.2922M} {440, 2922}

\bibitem[\protect\citeauthoryear{Mohammad \& Percival}{Mohammad \& Percival}{2022}]{Mohammad_2022}
Mohammad F.~G.,  Percival W.~J.,  2022, \mn@doi [\mnras] {10.1093/mnras/stac1458}, 514, 1289–1301

\bibitem[\protect\citeauthoryear{{Nadathur}}{{Nadathur}}{2016}]{Nadathur2016}
{Nadathur} S.,  2016, \mn@doi [\mnras] {10.1093/mnras/stw1340}, \href {http://adsabs.harvard.edu/abs/2016MNRAS.461..358N} {461, 358}

\bibitem[\protect\citeauthoryear{{Nadathur} \& {Crittenden}}{{Nadathur} \& {Crittenden}}{2016}]{nadathurcrit16}
{Nadathur} S.,  {Crittenden} R.,  2016, \mn@doi [\apjl] {10.3847/2041-8205/830/1/L19}, \href {https://ui.adsabs.harvard.edu/abs/2016ApJ...830L..19N} {830, L19}

\bibitem[\protect\citeauthoryear{{Nadathur} \& {Hotchkiss}}{{Nadathur} \& {Hotchkiss}}{2014}]{nadathur14_void_catalogue_paper}
{Nadathur} S.,  {Hotchkiss} S.,  2014, \mn@doi [\mnras] {10.1093/mnras/stu349}, \href {https://ui.adsabs.harvard.edu/abs/2014MNRAS.440.1248N} {440, 1248}

\bibitem[\protect\citeauthoryear{{Nadathur} \& {Hotchkiss}}{{Nadathur} \& {Hotchkiss}}{2015}]{nadathur15a}
{Nadathur} S.,  {Hotchkiss} S.,  2015, \mn@doi [\mnras] {10.1093/mnras/stv2131}, \href {https://ui.adsabs.harvard.edu/abs/2015MNRAS.454.2228N} {454, 2228}

\bibitem[\protect\citeauthoryear{{Nadathur} \& {Percival}}{{Nadathur} \& {Percival}}{2019}]{Nadathur:2019a}
{Nadathur} S.,  {Percival} W.~J.,  2019, \mn@doi [\mnras] {10.1093/mnras/sty3372}, \href {http://adsabs.harvard.edu/abs/2019MNRAS.483.3472N} {483, 3472}

\bibitem[\protect\citeauthoryear{{Nadathur}, {Hotchkiss}  \& {Crittenden}}{{Nadathur} et~al.}{2017a}]{Nadathur:2017a}
{Nadathur} S.,  {Hotchkiss} S.,   {Crittenden} R.,  2017a, \mn@doi [\mnras] {10.1093/mnras/stx336}, \href {http://adsabs.harvard.edu/abs/2017MNRAS.467.4067N} {467, 4067}

\bibitem[\protect\citeauthoryear{{Nadathur}, {Hotchkiss}  \& {Crittenden}}{{Nadathur} et~al.}{2017b}]{nadathur17}
{Nadathur} S.,  {Hotchkiss} S.,   {Crittenden} R.,  2017b, \mn@doi [\mnras] {10.1093/mnras/stx336}, \href {https://ui.adsabs.harvard.edu/abs/2017MNRAS.467.4067N} {467, 4067}

\bibitem[\protect\citeauthoryear{{Nadathur}, {Carter}, {Percival}, {Winther}  \& {Bautista}}{{Nadathur} et~al.}{2019a}]{Nadathur2019}
{Nadathur} S.,  {Carter} P.~M.,  {Percival} W.~J.,  {Winther} H.~A.,   {Bautista} J.~E.,  2019a, \mn@doi [\prd] {10.1103/PhysRevD.100.023504}, \href {https://ui.adsabs.harvard.edu/abs/2019PhRvD.100b3504N} {100, 023504}

\bibitem[\protect\citeauthoryear{{Nadathur}, {Carter}, {Percival}, {Winther}  \& {Bautista}}{{Nadathur} et~al.}{2019b}]{nadathur19}
{Nadathur} S.,  {Carter} P.~M.,  {Percival} W.~J.,  {Winther} H.~A.,   {Bautista} J.~E.,  2019b, \mn@doi [\prd] {10.1103/PhysRevD.100.023504}, \href {https://ui.adsabs.harvard.edu/abs/2019PhRvD.100b3504N} {100, 023504}

\bibitem[\protect\citeauthoryear{{Nadathur}, {Percival}, {Beutler}  \& {Winther}}{{Nadathur} et~al.}{2020a}]{nadathur20}
{Nadathur} S.,  {Percival} W.~J.,  {Beutler} F.,   {Winther} H.~A.,  2020a, \mn@doi [\prl] {10.1103/PhysRevLett.124.221301}, \href {https://ui.adsabs.harvard.edu/abs/2020PhRvL.124v1301N} {124, 221301}

\bibitem[\protect\citeauthoryear{{Nadathur} et~al.,}{{Nadathur} et~al.}{2020b}]{Nadathur2020}
{Nadathur} S.,  et~al., 2020b, \mn@doi [\mnras] {10.1093/mnras/staa3074}, \href {https://ui.adsabs.harvard.edu/abs/2020MNRAS.499.4140N} {499, 4140}

\bibitem[\protect\citeauthoryear{Neyrinck}{Neyrinck}{2008}]{neyrinck_zo}
Neyrinck M.~C.,  2008, \mn@doi [\mnras] {10.1111/j.1365-2966.2008.13180.x}, 386, 2101–2109

\bibitem[\protect\citeauthoryear{{Pandey} et~al.,}{{Pandey} et~al.}{2022}]{pandeyred}
{Pandey} S.,  et~al., 2022, \mn@doi [\prd] {10.1103/PhysRevD.106.043520}, \href {https://ui.adsabs.harvard.edu/abs/2022PhRvD.106d3520P} {106, 043520}

\bibitem[\protect\citeauthoryear{{Peebles}}{{Peebles}}{1980}]{peebles80}
{Peebles} P.~J.~E.,  1980, {The large-scale structure of the universe}.
{Princeton University Press}

\bibitem[\protect\citeauthoryear{Percival, Friedrich, Sellentin  \& Heavens}{Percival et~al.}{2021}]{Percival_2021}
Percival W.~J.,  Friedrich O.,  Sellentin E.,   Heavens A.,  2021, \mn@doi [\mnras] {10.1093/mnras/stab3540}, 510, 3207–3221

\bibitem[\protect\citeauthoryear{{Pisani} et~al.,}{{Pisani} et~al.}{2019}]{pisani19}
{Pisani} A.,  et~al., 2019, \baas, \href {https://ui.adsabs.harvard.edu/abs/2019BAAS...51c..40P} {51, 40}

\bibitem[\protect\citeauthoryear{{Planck Collaboration} et~al.,}{{Planck Collaboration} et~al.}{2016a}]{planck16_smica}
{Planck Collaboration} et~al., 2016a, \mn@doi [\aap] {10.1051/0004-6361/201525936}, \href {https://ui.adsabs.harvard.edu/abs/2016A&A...594A...9P} {594, A9}

\bibitem[\protect\citeauthoryear{{Planck Collaboration} et~al.,}{{Planck Collaboration} et~al.}{2016b}]{plancksz15}
{Planck Collaboration} et~al., 2016b, \mn@doi [\aap] {10.1051/0004-6361/201525833}, \href {http://adsabs.harvard.edu/abs/2016A%26A...594A..24P} {594, A24}

\bibitem[\protect\citeauthoryear{{Planck Collaboration} et~al.,}{{Planck Collaboration} et~al.}{2020a}]{Planck2018_cosmo}
{Planck Collaboration} et~al., 2020a, \mn@doi [\aap] {10.1051/0004-6361/201833910}, \href {https://ui.adsabs.harvard.edu/abs/2020A&A...641A...6P} {641, A6}

\bibitem[\protect\citeauthoryear{{Planck Collaboration} et~al.,}{{Planck Collaboration} et~al.}{2020b}]{planck18_lensing}
{Planck Collaboration} et~al., 2020b, \mn@doi [\aap] {10.1051/0004-6361/201833886}, \href {https://ui.adsabs.harvard.edu/abs/2020A&A...641A...8P} {641, A8}

\bibitem[\protect\citeauthoryear{Porredon et~al.,}{Porredon et~al.}{2022}]{Porredon_2022}
Porredon A.,  et~al., 2022, \mn@doi [Phys. Rev. D] {10.1103/physrevd.106.103530}, 106

\bibitem[\protect\citeauthoryear{{Radinovi{\'c}} et~al.,}{{Radinovi{\'c}} et~al.}{2023}]{radinovicvoxel}
{Radinovi{\'c}} S.,  et~al., 2023, \mn@doi [\aap] {10.1051/0004-6361/202346121}, \href {https://ui.adsabs.harvard.edu/abs/2023A&A...677A..78R} {677, A78}

\bibitem[\protect\citeauthoryear{{Raghunathan} et~al.,}{{Raghunathan} et~al.}{2019}]{raghunathan19}
{Raghunathan} S.,  et~al., 2019, \mn@doi [\apj] {10.3847/1538-4357/ab01ca}, \href {https://ui.adsabs.harvard.edu/\#abs/2019ApJ...872..170R} {872, 170}

\bibitem[\protect\citeauthoryear{{Raghunathan}, {Nadathur}, {Sherwin}  \& {Whitehorn}}{{Raghunathan} et~al.}{2020}]{Raghunathan2019}
{Raghunathan} S.,  {Nadathur} S.,  {Sherwin} B.~D.,   {Whitehorn} N.,  2020, \mn@doi [\apj] {10.3847/1538-4357/ab6f05}, \href {https://ui.adsabs.harvard.edu/abs/2020ApJ...890..168R} {890, 168}

\bibitem[\protect\citeauthoryear{{Riess}, {Casertano}, {Yuan}, {Macri}  \& {Scolnic}}{{Riess} et~al.}{2019}]{Riess2019}
{Riess} A.~G.,  {Casertano} S.,  {Yuan} W.,  {Macri} L.~M.,   {Scolnic} D.,  2019, \mn@doi [\apj] {10.3847/1538-4357/ab1422}, \href {https://ui.adsabs.harvard.edu/abs/2019ApJ...876...85R} {876, 85}

\bibitem[\protect\citeauthoryear{{Rozo}, {Rykoff}, {Abate}  \& et al.}{{Rozo} et~al.}{2016a}]{Rozo2015}
{Rozo} E.,  {Rykoff} E.~S.,  {Abate} A.,   et al. 2016a, \mn@doi [\mnras] {10.1093/mnras/stw1281}, \href {http://adsabs.harvard.edu/abs/2016MNRAS.461.1431R} {461, 1431}

\bibitem[\protect\citeauthoryear{{Rozo} et~al.,}{{Rozo} et~al.}{2016b}]{redmagic}
{Rozo} E.,  et~al., 2016b, \mn@doi [\mnras] {10.1093/mnras/stw1281}, \href {http://adsabs.harvard.edu/abs/2016MNRAS.461.1431R} {461, 1431}

\bibitem[\protect\citeauthoryear{{Rykoff}, {Rozo}, {Busha}  \& et al.}{{Rykoff} et~al.}{2014}]{Rykoff2014}
{Rykoff} E.~S.,  {Rozo} E.,  {Busha} M.~T.,   et al. 2014, \mn@doi [\apj] {10.1088/0004-637X/785/2/104}, \href {http://adsabs.harvard.edu/abs/2014ApJ...785..104R} {785, 104}

\bibitem[\protect\citeauthoryear{{Sachs} \& {Wolfe}}{{Sachs} \& {Wolfe}}{1967}]{Sachs1967}
{Sachs} R.~K.,  {Wolfe} A.~M.,  1967, \mn@doi [ApJL] {10.1086/148982}, \href {http://adsabs.harvard.edu/abs/1967ApJ...147...73S} {147, 73}

\bibitem[\protect\citeauthoryear{{Sch{\"a}fer}, {Pfrommer}, {Hell}  \& {Bartelmann}}{{Sch{\"a}fer} et~al.}{2006}]{Schaefer:2006}
{Sch{\"a}fer} B.~M.,  {Pfrommer} C.,  {Hell} R.~M.,   {Bartelmann} M.,  2006, \mn@doi [\mnras] {10.1111/j.1365-2966.2006.10622.x}, \href {http://adsabs.harvard.edu/abs/2006MNRAS.370.1713S} {370, 1713}

\bibitem[\protect\citeauthoryear{{Schmittfull} \& {Seljak}}{{Schmittfull} \& {Seljak}}{2018}]{schmittfull18}
{Schmittfull} M.,  {Seljak} U.,  2018, \mn@doi [\prd] {10.1103/PhysRevD.97.123540}, \href {https://ui.adsabs.harvard.edu/abs/2018PhRvD..97l3540S} {97, 123540}

\bibitem[\protect\citeauthoryear{Sheth \& van~de Weygaert}{Sheth \& van~de Weygaert}{2004}]{Sheth:2004}
Sheth R.~K.,  van~de Weygaert R.,  2004, \mn@doi [\mnras] {10.1111/j.1365-2966.2004.07661.x}, 350, 517

\bibitem[\protect\citeauthoryear{{Springel}}{{Springel}}{2005}]{springel2005}
{Springel} V.,  2005, \mn@doi [\mnras] {10.1111/j.1365-2966.2005.09655.x}, \href {http://adsabs.harvard.edu/abs/2005MNRAS.364.1105S} {364, 1105}

\bibitem[\protect\citeauthoryear{Stein, Alvarez, Bond, van Engelen  \& Battaglia}{Stein et~al.}{2020}]{webskysim}
Stein G.,  Alvarez M.~A.,  Bond J.~R.,  van Engelen A.,   Battaglia N.,  2020, \mn@doi [Journal of Cosmology and Astroparticle Physics] {10.1088/1475-7516/2020/10/012}, 2020, 012

\bibitem[\protect\citeauthoryear{{Sutter} et~al.,}{{Sutter} et~al.}{2015}]{VIDE:2015}
{Sutter} P.~M.,  et~al., 2015, \mn@doi [Astronomy and Computing] {10.1016/j.ascom.2014.10.002}, \href {http://adsabs.harvard.edu/abs/2015A%26C.....9....1S} {9, 1}

\bibitem[\protect\citeauthoryear{Sánchez et~al.,}{Sánchez et~al.}{2016}]{20sanchez2d}
Sánchez C.,  et~al., 2016, \mn@doi [\mnras] {10.1093/mnras/stw2745}, 465, 746–759

\bibitem[\protect\citeauthoryear{{Tanimura}, {Aghanim}, {Bonjean}, {Malavasi}  \& {Douspis}}{{Tanimura} et~al.}{2020}]{tanimura}
{Tanimura} H.,  {Aghanim} N.,  {Bonjean} V.,  {Malavasi} N.,   {Douspis} M.,  2020, \mn@doi [\aap] {10.1051/0004-6361/201937158}, \href {https://ui.adsabs.harvard.edu/abs/2020A&A...637A..41T} {637, A41}

\bibitem[\protect\citeauthoryear{{Vainshtein}}{{Vainshtein}}{1972}]{vain1972}
{Vainshtein} A.~I.,  1972, \mn@doi [Physics Letters B] {10.1016/0370-2693(72)90147-5}, \href {https://ui.adsabs.harvard.edu/abs/1972PhLB...39..393V} {39, 393}

\bibitem[\protect\citeauthoryear{Verde, Treu  \& Riess}{Verde et~al.}{2019}]{verde2019}
Verde L.,  Treu T.,   Riess A.~G.,  2019, \mn@doi [Nature Astronomy] {10.1038/s41550-019-0902-0}, 3, 891

\bibitem[\protect\citeauthoryear{Vielzeuf et~al.,}{Vielzeuf et~al.}{2020}]{vielzeuf19}
Vielzeuf P.,  et~al., 2020, \mn@doi [\mnras] {10.1093/mnras/staa3231}, 500, 464

\bibitem[\protect\citeauthoryear{{Vielzeuf} et~al.,}{{Vielzeuf} et~al.}{2021}]{vielzeufy1}
{Vielzeuf} P.,  et~al., 2021, \mn@doi [\mnras] {10.1093/mnras/staa3231}, \href {https://ui.adsabs.harvard.edu/abs/2021MNRAS.500..464V} {500, 464}

\bibitem[\protect\citeauthoryear{Vielzeuf, Calabrese, Carbone, Fabbian  \& Baccigalupi}{Vielzeuf et~al.}{2023}]{vielzeuf23}
Vielzeuf P.,  Calabrese M.,  Carbone C.,  Fabbian G.,   Baccigalupi C.,  2023, \mn@doi [Journal of Cosmology and Astroparticle Physics] {10.1088/1475-7516/2023/08/010}, 2023, 010

\bibitem[\protect\citeauthoryear{Villaescusa-Navarro, Vogelsberger, Viel  \& Loeb}{Villaescusa-Navarro et~al.}{2013}]{villa2013}
Villaescusa-Navarro F.,  Vogelsberger M.,  Viel M.,   Loeb A.,  2013, \mn@doi [\mnras] {10.1093/mnras/stt452}, 431, 3670

\bibitem[\protect\citeauthoryear{Woodfinden, Nadathur, Percival, Radinovic, Massara  \& Winther}{Woodfinden et~al.}{2022}]{woodfinden2022}
Woodfinden A.,  Nadathur S.,  Percival W.~J.,  Radinovic S.,  Massara E.,   Winther H.~A.,  2022, \mn@doi [\mnras] {10.1093/mnras/stac2475}, 516, 4307

\bibitem[\protect\citeauthoryear{Woodfinden, Percival, Nadathur, Winther, Fraser, Massara, Paillas  \& Radinovi{\'{c} }}{Woodfinden et~al.}{2023}]{Woodfinden_2023}
Woodfinden A.,  Percival W.~J.,  Nadathur S.,  Winther H.~A.,  Fraser T.~S.,  Massara E.,  Paillas E.,   Radinovi{\'{c} } S.,  2023, \mn@doi [\mnras] {10.1093/mnras/stad1725}, 523, 6360

\makeatother
\end{thebibliography}
%%%%%%%%%%%%%%%%%%%%%%%%%%%%%%%%%%%%%%%%%%%%%%%%%%
\appendix
\section{ADDITIONAL PLOTS}
\label{sec:Appendix}
\begin{figure*}
    \begin{minipage}{\textwidth}
    \centering
    \includegraphics[width=0.95\textwidth]{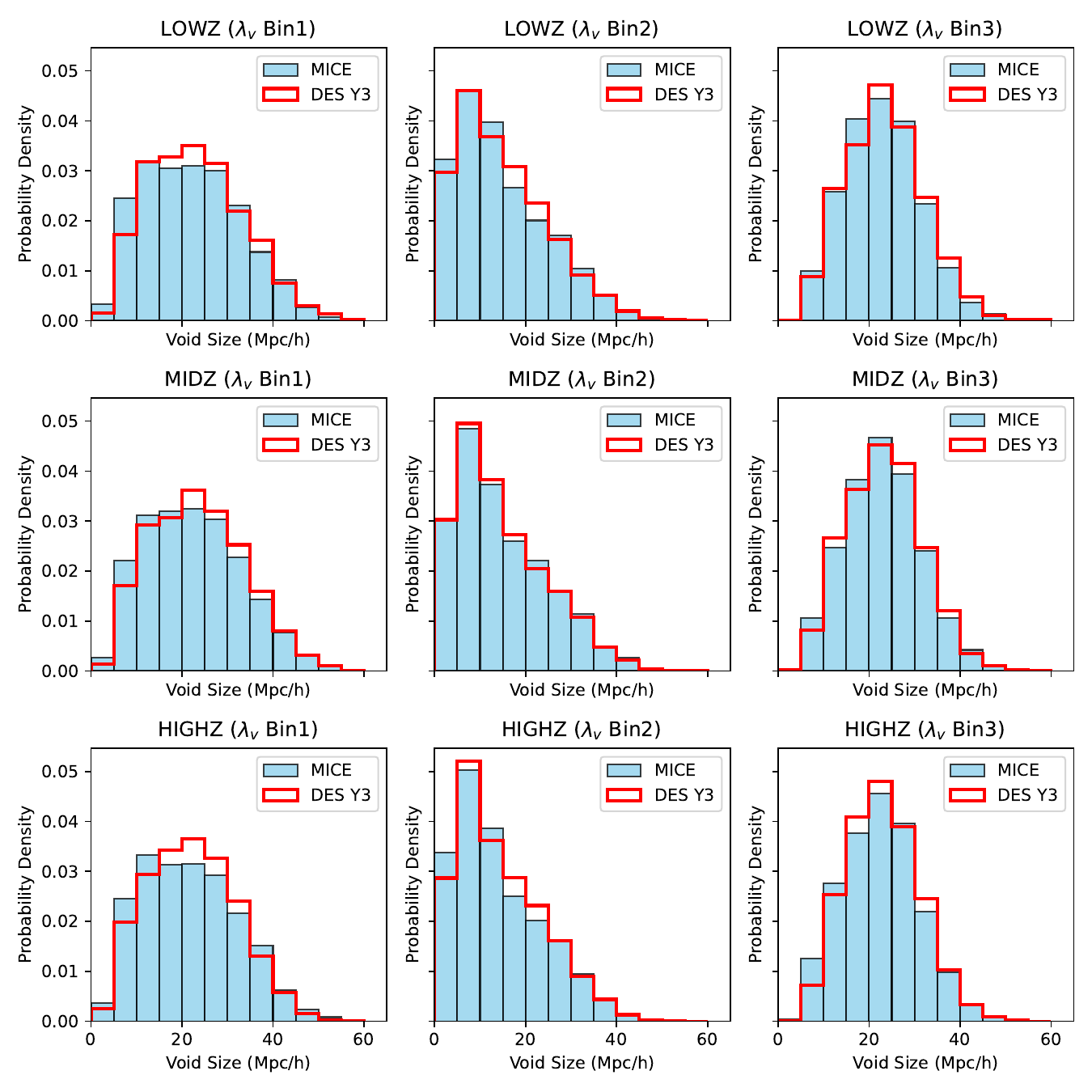}
    \caption{This plot illustrates the \texttt{Voxel} void size distributions for both MICE and DESY3 using a normalized histogram, segmented across 9 redshift bins. Notably, the number of voids inreases with increasing redshift bin. Additionally, voids within the same \(\lambda_v\) bins, yet across differing redshift bins, exhibit a consistent size distribution pattern. This consistency emphasizes the functional relationship between \(\lambda_v\) and \(R_v\), as shown in Equation \ref{eq:lambda_v}.}
    \label{fig:voxel_size_distributions}
    \end{minipage}
\end{figure*}

\begin{figure*}
    \centering % This centers the figure in the text.
    \includegraphics[width=0.7\textwidth]{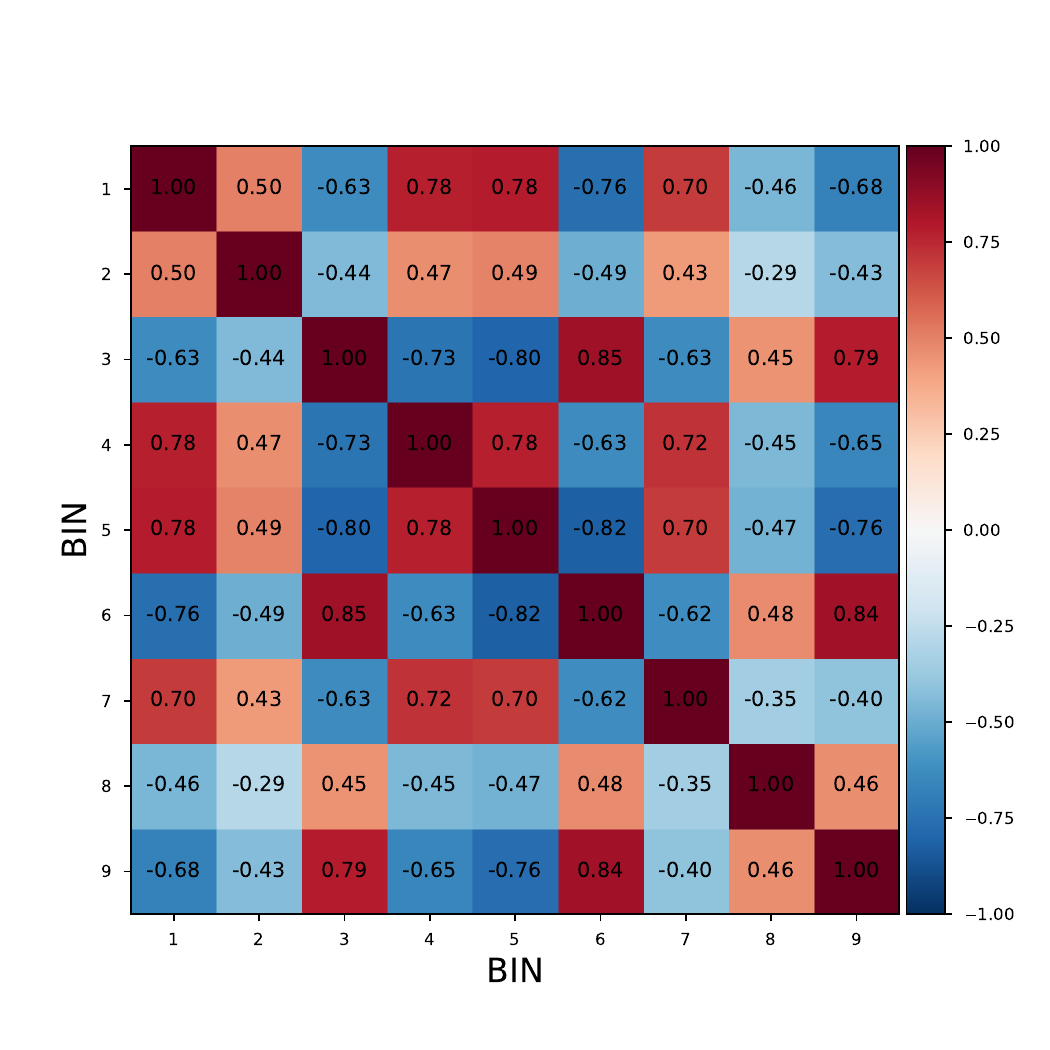} % This scales the figure to the text width, making it larger.
    \caption{This figure shows the correlation matrix for 9 \texttt{Voxel} bins. The bins range from BIN 1, representing LOWZ \(\lambda_v\) Bin 1, to BIN 9, representing HIGHZ \(\lambda_v\) Bin 3.}
    \label{fig:voxel_correlation_matrix}
\end{figure*}

We show additional plots that can be useful for our analysis. See Figure \ref{fig:voxel_size_distributions} for the size distribution of \texttt{Voxel} voids and Figure \ref{fig:voxel_correlation_matrix} for their correlation matrix.

%\appendix
\section{Author Affiliations}
\label{sec:affiliations}
{\small
$^{1}$ Institut de F\'{\i}sica d'Altes Energies (IFAE), The Barcelona Institute of Science and Technology, Campus UAB, 08193 Bellaterra (Barcelona) Spain\\
$^{2}$ Institute of Cosmology and Gravitation, University of Portsmouth, Portsmouth, PO1 3FX, UK\\
$^{3}$ Institute of Theoretical Astrophysics, University of Oslo. P.O. Box 1029 Blindern, NO-0315 Oslo, Norway\\
$^{4}$ Institut d'Estudis Espacials de Catalunya (IEEC), 08034 Barcelona, Spain\\
$^{5}$ MTA-CSFK Lend\"ulet "Momentum" Large-Scale Structure (LSS) Research Group, 1121 Budapest, Konkoly Thege Mikl\'os \'ut 15-17, Hungary \\
$^{6}$ Instituci\'o Catalana de Recerca i Estudis Avan\c{c}ats, E-08010 Barcelona, Spain\\
$^{7}$ University Observatory, Faculty of Physics, Ludwig-Maximilians-Universit\"at, Scheinerstr. 1, 81679 Munich, Germany\\
$^{8}$ Department of Physics and Astronomy, University of Pennsylvania, Philadelphia, PA 19104, USA\\
$^{9}$ Center for Astrophysical Surveys, National Center for Supercomputing Applications, 1205 West Clark St., Urbana, IL 61801, USA\\
$^{10}$ Physics Department, 2320 Chamberlin Hall, University of Wisconsin-Madison, 1150 University Avenue Madison, WI 53706-1390\\
$^{11}$ Department of Astronomy and Astrophysics, University of Chicago, Chicago, IL 60637, USA\\
$^{12}$ Center for Astrophysical Surveys, National Center for Supercomputing Applications, 1205 West Clark St., Urbana, IL 61801, USA\\
$^{13}$ Department of Astronomy, University of Geneva, ch. d'\'Ecogia 16, CH-1290 Versoix, Switzerland\\
$^{14}$ Laborat\'orio Interinstitucional de e-Astronomia - LIneA, Rua Gal. Jos\'e Cristino 77, Rio de Janeiro, RJ - 20921-400, Brazil\\
$^{15}$ Center for Cosmology and Astro-Particle Physics, The Ohio State University, Columbus, OH 43210, USA\\
$^{16}$ Kavli Institute for Particle Astrophysics \& Cosmology, P. O. Box 2450, Stanford University, Stanford, CA 94305, USA\\
$^{17}$ Brookhaven National Laboratory, Bldg 510, Upton, NY 11973, USA\\
$^{18}$ Fermi National Accelerator Laboratory, P. O. Box 500, Batavia, IL 60510, USA\\
$^{19}$ Cerro Tololo Inter-American Observatory, NSF's National Optical-Infrared Astronomy Research Laboratory, Casilla 603, La Serena, Chile\\
$^{20}$ Laborat\'orio Interinstitucional de e-Astronomia - LIneA, Rua Gal. Jos\'e Cristino 77, Rio de Janeiro, RJ - 20921-400, Brazil\\
$^{21}$ Fermi National Accelerator Laboratory, P. O. Box 500, Batavia, IL 60510, USA\\
$^{22}$ Department of Physics, University of Michigan, Ann Arbor, MI 48109, USA\\
$^{23}$ Institute of Cosmology and Gravitation, University of Portsmouth, Portsmouth, PO1 3FX, UK\\
$^{24}$ CNRS, UMR 7095, Institut d'Astrophysique de Paris, F-75014, Paris, France\\
$^{25}$ Sorbonne Universit\'es, UPMC Univ Paris 06, UMR 7095, Institut d'Astrophysique de Paris, F-75014, Paris, France\\
$^{26}$ University Observatory, Faculty of Physics, Ludwig-Maximilians-Universit\"at, Scheinerstr. 1, 81679 Munich, Germany\\
$^{27}$ Department of Physics \& Astronomy, University College London, Gower Street, London, WC1E 6BT, UK\\
$^{28}$ Instituto de Astrofisica de Canarias, E-38205 La Laguna, Tenerife, Spain\\
$^{29}$ Institut de F\'{\i}sica d'Altes Energies (IFAE), The Barcelona Institute of Science and Technology, Campus UAB, 08193 Bellaterra (Barcelona) Spain\\
$^{30}$ Physics Department, William Jewell College, Liberty, MO, 64068\\
$^{31}$ Hamburger Sternwarte, Universit\"{a}t Hamburg, Gojenbergsweg 112, 21029 Hamburg, Germany\\
$^{32}$ Centro de Investigaciones Energ\'eticas, Medioambientales y Tecnol\'ogicas (CIEMAT), Madrid, Spain\\
$^{33}$ Department of Physics, IIT Hyderabad, Kandi, Telangana 502285, India\\
$^{34}$ Jet Propulsion Laboratory, California Institute of Technology, 4800 Oak Grove Dr., Pasadena, CA 91109, USA\\
$^{35}$ Department of Physics and Astronomy, University of Pennsylvania, Philadelphia, PA 19104, USA\\
$^{36}$ School of Mathematics and Physics, University of Queensland, Brisbane, QLD 4072, Australia\\
$^{37}$ Santa Cruz Institute for Particle Physics, Santa Cruz, CA 95064, USA\\
$^{38}$ Center for Astrophysics | Harvard \& Smithsonian, 60 Garden Street, Cambridge, MA 02138, USA\\
$^{39}$ Australian Astronomical Optics, Macquarie University, North Ryde, NSW 2113, Australia\\
$^{40}$ Lowell Observatory, 1400 Mars Hill Rd, Flagstaff, AZ 86001, USA\\
$^{41}$ George P. and Cynthia Woods Mitchell Institute for Fundamental Physics and Astronomy, and Department of Physics and Astronomy, Texas A\&M University, College Station, TX 77843, USA\\
$^{42}$ LPSC Grenoble - 53, Avenue des Martyrs 38026 Grenoble, France\\
$^{43}$ Max Planck Institute for Extraterrestrial Physics, Giessenbachstrasse, 85748 Garching, Germany\\
$^{44}$ Department of Astrophysical Sciences, Princeton University, Peyton Hall, Princeton, NJ 08544, USA\\
$^{45}$ Kavli Institute for Cosmological Physics, University of Chicago, Chicago, IL 60637, USA\\
$^{46}$ Kavli Institute for Particle Astrophysics \& Cosmology, P. O. Box 2450, Stanford University, Stanford, CA 94305, USA\\
$^{47}$ SLAC National Accelerator Laboratory, Menlo Park, CA 94025, USA\\
$^{48}$ School of Physics and Astronomy, University of Southampton, Southampton, SO17 1BJ, UK\\
$^{49}$ Observat\'orio Nacional, Rua Gal. Jos\'e Cristino 77, Rio de Janeiro, RJ - 20921-400, Brazil\\
$^{50}$ Computer Science and Mathematics Division, Oak Ridge National Laboratory, Oak Ridge, TN 37831, USA\\
$^{51}$ Department of Astronomy, University of California, Berkeley, 501 Campbell Hall, Berkeley, CA 94720, USA\\
$^{52}$ Lawrence Berkeley National Laboratory, 1 Cyclotron Road, Berkeley, CA 94720, USA\\
$^{53}$ Universit\"ats-Sternwarte, Fakult\"at f\"ur Physik, Ludwig-Maximilians Universit\"at M\"unchen, Scheinerstr. 1, 81679 M\"unchen, Germany\\
$^{54}$ Konkoly Observatory, HUN-REN CSFK, MTA Centre of Excellence, Budapest, Konkoly Thege Mikl\'os {\'u}t 15-17. H-1121 Hungary \\
}

%\include{DATASETS SIMULATION}
%\include{METHOD}
%\include{RESULTS}
%\include{Conclusions}
%\include{Appendix}
%%%%%%%%%%%%%%%%%%%%%%%%%%%%%%%%%%%%%%%%%%%%
%%%%%%%%%%%%%%%%%%%%%%%%%%%%%%%%%%%%%%%%%%%%
% Don't change these lines
%\bsp	% typesetting comment
\AtEndDocument{
  %\clearpage % Clears the current page
  %\thispagestyle{empty} % Sets the page style to empty
  %\null % Puts nothing in the text block
  %\clearpage % Ensures any automatic insertion after this point does not affect the output
}

\end{document}